\PassOptionsToPackage{table,xcdraw}{xcolor}
\documentclass[sigconf]{acmart}

\usepackage{balance}

\usepackage[english]{babel}
\usepackage{blindtext}
\usepackage{tikz}

\usepackage{paralist}
\usepackage{soul}
\usepackage{placeins}

\newcommand{\eat}[1]{}

\usepackage[absolute,showboxes]{textpos} %

\usepackage{tabularx}
\newcolumntype{L}[1]{>{\raggedright\arraybackslash}p{#1}}
\newcolumntype{C}[1]{>{\centering\arraybackslash}p{#1}}
\newcolumntype{R}[1]{>{\raggedleft\arraybackslash}p{#1}}
\newcommand{\provider}{provider\xspace}
\newcommand{\providers}{providers\xspace}

\newcommand{\iotbackendprovider}{IoT backend provider\xspace}
\newcommand{\iotbackendproviders}{IoT backend providers\xspace}

\usepackage{enumitem}
\setlist{nolistsep}
\usepackage{graphicx}
\usepackage[export]{adjustbox}
\usepackage{epstopdf}
\usepackage{grffile}
\usepackage{amsmath}
\usepackage{multicol,multirow}
\usepackage{rotating,url}
\usepackage{enumitem}
\usepackage{color}
\usepackage{subcaption}
\usepackage{xspace}
\usepackage{ifpdf}
\usepackage{mdwlist}
\usepackage{colortbl}
\usepackage{caption}
\usepackage{booktabs}
\usepackage{multirow}
\usepackage{enumitem}
\usepackage{xfrac}
\usepackage{afterpage}%
\usepackage{pdflscape}
\usepackage{longtable}
\usepackage{hhline}
\usepackage{lscape}
\usepackage{cleveref}

\newlength{\oldtextfloatsep}
\newcommand{\ie}{i.e., \@}

\usepackage{enumitem}
\usepackage{mdwlist}
\setlist{nolistsep}
\setlist[description]{noitemsep,topsep=0pt,parsep=0pt,partopsep=0pt,leftmargin=0pt}
\usepackage{pifont}%

\usepackage{everypage}

\newcommand{\tone}{T1\xspace}%
\newcommand{\ttwo}{T2\xspace}%
\newcommand{\ofive}{O5\xspace}%
\newcommand{\othree}{O3\xspace}%
\newcommand{\tfour}{T4\xspace}%
\newcommand{\tthree}{T3\xspace}%

\newcommand{\done}{D1\xspace}%

\newcommand{\dthree}{D3\xspace}%
\newcommand{\dfour}{D4\xspace}%

\newcommand{\dsix}{D6\xspace}%

\newcommand{\frontend}{frontend\xspace}%
\copyrightyear{2022} 
\acmYear{2022} 
\setcopyright{rightsretained} 
\acmConference[IMC '22]{Proceedings of the 22nd ACM Internet Measurement Conference}{October 25--27, 2022}{Nice, France}
\acmBooktitle{Proceedings of the 22nd ACM Internet Measurement Conference (IMC '22), October 25--27, 2022, Nice, France}
\acmDOI{10.1145/3517745.3561431}
\acmISBN{978-1-4503-9259-4/22/10}

\long\def\comment#1{}

\begin{document}
\clubpenalty=10000 
\widowpenalty = 10000

\title[Deep Dive into the IoT Backend Ecosystem]{Deep Dive into the IoT Backend Ecosystem}

\author{Said Jawad Saidi}
\affiliation{%
  \institution{Max Planck Institute for Informatics, Saarland University}
  \country{}
}
\author{Srdjan Matic}
\affiliation{%
  \institution{IMDEA Software Institute}
  \country{}
}
\author{Oliver Gasser}
\affiliation{%
  \institution{Max Planck Institute for Informatics}
  \country{}
}
\author{Georgios Smaragdakis}
\affiliation{%
  \institution{TU Delft}
  \country{}
}
\author{Anja Feldmann}
\affiliation{%
  \institution{Max Planck Institute for Informatics}
  \country{}
}

\settopmatter{printfolios=true}

\setlength{\TPHorizModule}{\paperwidth}
\setlength{\TPVertModule}{\paperheight}
\TPMargin{5pt}
\begin{textblock}{0.8}(0.1,0.02)
     \noindent
     \footnotesize
     If you cite this paper, please use the ACM IMC reference:
     Said Jawad Saidi, Srdjan Matic, Oliver Gasser, Georgios Smaragdakis, and Anja Feldmann. 2022.
     Deep Dive into the IoT Backend Ecosystem. In \textit{Proceedings of the 22nd ACM Internet Measurement Conference (IMC ’22), October 25–27, 2022, Nice, France}. ACM, New York, NY, USA, 16 pages. \url{https://doi.org/10.1145/3517745.3561431}
\end{textblock}
 
\begin{abstract}
Internet of Things (IoT) devices are becoming increasingly ubiquitous, e.g., at
home, in enterprise environments, and in production lines. To support the
advanced functionalities of IoT devices, IoT vendors as well as service and
cloud companies operate IoT backends---the focus of this paper. We propose a
methodology to identify and locate them by (a) compiling a list of domains used
exclusively by major IoT backend providers and (b) then identifying their server
IP addresses. We rely on multiple sources, including IoT backend provider
documentation, passive DNS data, and active scanning. For analyzing IoT traffic
patterns, we rely on passive network flows from a major European ISP.

Our analysis focuses on the top IoT backends and unveils
diverse operational strategies---from operating their own infrastructure to
utilizing the public cloud. We find that the majority of the top IoT backend
providers are located in multiple locations and countries. Still, a handful are
located only in one country, which could raise regulatory scrutiny as the
client IoT devices are located in other regions. Indeed, our analysis shows that
up to 35\% of IoT traffic is exchanged with IoT backend servers located in other
continents. We also find that at least six of the top IoT backends rely on
other IoT backend providers. We also evaluate if cascading effects among the IoT
backend providers are possible in the event of an outage, a misconfiguration, or
an attack.
 \end{abstract}



\maketitle

\section{Introduction}\label{sec:intro} 

Internet of Things (IoT) devices are increasingly deployed at home, office,
retails, and production lines to enable rich and complex applications,
including smart home, video surveillance, voice assistance, content
recommendation, and logistics, to just name a few. Many of these applications rely on functionalities that cannot be fully deployed on 
IoT devices directly. 

Given that IoT devices simply lack the required computing, memory, and energy
resources for computationally demanding applications or additional data, it
is common to offload part of the application to a backend in the ``cloud''. For
example, applications that rely on machine learning 
are often easier to operate in the cloud, which is computationally more powerful
and has access to readily available machine learning libraries~\cite{google-AI,Amazon-AI},
rather than on the IoT device itself. 
A low-cost IoT camera typically streams its video to the cloud, where the main
computation takes place, e.g., to identify suspicious activity and trigger an
alarm in real-time. Moreover, many companies that use IoT devices commercially,
e.g., within a production line or for logistics, collect all data in the cloud
for analytics and operational decisions~\cite{IoT-production}. 
Thus, these clouds act as the {\it backend} of IoT applications.

IoT devices also lack the storage required, e.g., for content-centric
applications. Thus, such IoT devices need IoT backend servers to download or
upload content required by the application. For example, content
recommendations require the user's profile and have to merge it against
the available content~\cite{amazon-personalization},
which may not be possible on the IoT device itself.
Moreover, IoT device security and functionality often depend on an IoT backend.
One prominent example is software updates---many IoT devices periodically check
if software updates are available. Other IoT vendors or application
providers push notifications to the IoTs when such updates are available.

As the number of deployed IoTs and their functionality increases rapidly, the
demands for the IoT backend---in terms of capabilities and traffic---increase as well.
During the last years, we have observed a shift toward building special-purpose
clouds to support IoT applications and cope with the increasing demand.
Recently, big technology giants, such as Amazon~\cite{awsiot},
Google~\cite{googleiot}, and Microsoft~\cite{azureiot} started to offer IoT backend
solutions as-a-service. Such companies are IoT backend providers and
enable third-party IoT application providers to scale up and deliver their solutions to
potentially billions of IoTs deployed around the globe.

Despite the critical role that these IoT backend providers play in the operation and
security of IoT applications~\cite{alrawi2019sok, He-iot-cloud,Zhou-iot-cloud,Jia-iot-cloud}, little is known about their locations, strategies,
and volume share. Indeed, much of the work in the IoT area has focused on the
inference of IoT clients~\cite{IMC2020-IoT,iotfinder,IoT-All-Things-2019} or
general-purpose cloud providers or content delivery
networks~\cite{cloud-providers2020,cloud-connectivity,SIGCOMM2021-hypergiants} that may
also support IoT services. In this paper, we turn our attention to the IoT
backend providers. We develop new methods to identify their footprints 
and gain insights into their modus operandi.

We also investigate if IoTs and IoT backend servers are in the same
geographic location and jurisdiction. Indeed, data sovereignty and protecting private user data leaked by IoTs are at the heart
of the current debate. The European Union General Data
Protection Regulation (GDPR)~\cite{EU-GDPR} was put into effect on May 25, 2018
to protect user privacy and regulate the transfer of personal data only under strict
conditions and with user consent. The GDPR levies fines against those who violate users' privacy and security
standards, with penalties reaching twenty million euros or up to 4\% of the
annual worldwide turnover of the preceding financial year in the case of an
enterprise, whichever is greater. Moreover, the EU is currently working on a new
regulation targeting smart devices, e.g., IoTs, with cybersecurity and
privacy
risks~\cite{Reuters-IoT-cybersecurity,EU-IoT-cybersecurity,EU-device-sec-2022}. 
Thus, it is important to better understand the
interaction of IoT devices with remote backend providers. 

\noindent Our contributions can be summarized as follows: 

\begin{itemize} [leftmargin=*]

\item We develop a methodology to infer the network and physical location of
  major IoT backend providers. Our methodology relies on a fusion of
  information from public documentation, passive DNS, and active
  measurements.%

\item We analyze the IoT backend ecosystem with regard to deployment,
operation, and dependencies. While most popular IoT backend providers have
footprints that cover multiple locations and countries, our analysis shows that
some of them operate only in one country or rely on infrastructure from other
IoT backend providers.

\item Our analysis shows that it is not unusual for IoT protocols, e.g., MQTT, to
use non-standard ports or reuse Web ports. The latter makes the identification
of IoT backend infrastructure as well as IoT traffic challenging using
traditional methods---our proposed methodology resolves this issue.

\item Using passive data from a major European ISP, we examine the IoT traffic
patterns of multiple providers at scale. We notice that a substantial fraction
(around 35\%) of IoT traffic is exchanged with IoT backend servers outside Europe, which raises
both performance and regulatory concerns.
 
\item Our traffic analysis highlights that both
the IoT population and activity per application differ vastly. While some
applications behave more like the typical user-generated traffic, i.e., diurnal
patterns, peak evening hours, and downstream-heavy; this is not the case for
all IoT applications.

\item We comment on shortcomings of the current IoT backend ecosystem 
  and assess the impact of a large-scale outage in one of the major IoT backend
providers on IoT connectivity to backend servers.

\item To enable follow-up research in the area, we make the tools, scripts, and artifacts for extracting the regular expressions and IoT backend domains publicly
  available~\cite{iot-backend-github}.

\end{itemize}

\vspace{1em}
\noindent {\bf Scope of the paper:}
Our study is curiosity-driven, and we try to understand the evolving IoT backend
ecosystem to inform future studies by computer scientists, economists,
and policymakers. As we are not aware of the companies' business strategies, we do not take a position regarding their deployment decisions and
operation. Rather, we characterize the current state of the IoT ecosystem.
This study is not a head-to-head comparison of different and possibly competing
IoT companies.

\section{Scenario and related work}\label{sec:background} %

\begin{figure} [t]
	\captionsetup{skip=.25em,font=small}
	\includegraphics[width=1\linewidth,valign=t]{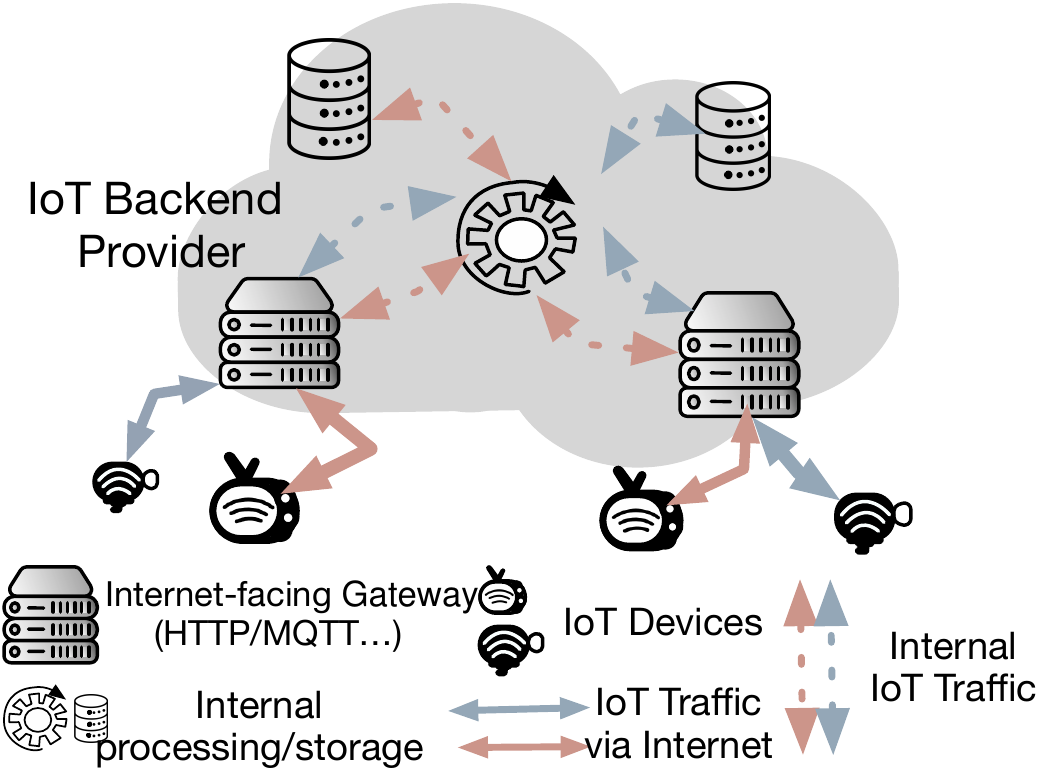}
	\caption{IoT backend provider architecture.}
	\label{fig:iot-cloud-platform}
\end{figure}

We first describe the setting of our measurement study and introduce
terminology. Then, we summarize related work.

\subsection{IoT Backend Providers}

Today, many of the IoT
vendors~\cite{boschiot,fujitsuiot,siemensmindsphere,huaweiiot}, technology
giants~\cite{googleiot,ibmiot,oracleiot,sapiot,alibabaiot,baidoiot}, and cloud
providers~\cite{awsiot,azureiot} offer sophisticated IoT platform
solutions. These solutions allow developers to deploy new services, support
existing applications, collect data, or remotely manage and configure IoT
devices. Typically, these IoT platforms have three major components: $(i)$
software/hardware on IoT devices, $(ii)$ Internet-facing gateway servers, and
$(iii)$ the internal storage/processing systems, e.g., for machine
learning. Figure~\ref{fig:iot-cloud-platform} depicts the main components of a
generic IoT platform. 

In this paper, we identify and characterize the public part of IoT platforms,
i.e., the Internet-facing gateway servers, that we refer to as {\it IoT backend}, see
Figure~\ref{fig:iot-cloud-platform}. IoT
backends facilitate the data exchange between IoT devices, internal
systems, and possibly other platforms. We refer to companies that operate such
gateway infrastructures as {\it IoT backend providers}.

Note that IoT platforms are sophisticated entities. Thus, our study focuses on their public IPs---the gateways that enable the exchange and flow of data between the IoT devices and the internal systems of the IoT platforms.
Thus, the following aspects are out of scope: $(i)$ the software and hardware installed on IoT devices, 
$(ii)$ internal processing and storage systems of IoT platforms, in particular, since these
are typically not publicly accessible, and $(iii)$ private interconnections between
cloud providers and IoT platforms~\cite{cloud-hiding}.

In this paper, we consider devices that contact IoT backends, which are
dedicated to supporting functionalities of IoT devices, as IoT devices. IoT
devices range from smart meters at home to smart TVs, voice assistants, and
logistics monitors. What they have in common is that they contact IoT backend
providers that support their functionalities.

\subsection{Related Work}

Most of the related work in this area focused on identifying either IoT devices
themselves or their vendors in the wild rather than the IoT backend.\\

\vspace{-0.75em}
\noindent{\bf Instrumented Testbeds.}
Previous
work~\cite{IoTInspector,PETS20_DuboisKMPCH20,Information-Exposure-IMC2019,haddadi2018siotome,alrawi2019sok}
uses sophisticated testbeds or home environments to collect full packet captures
to generate IoT device signatures. Other studies~\cite{MUD-sigcomm2018w} use
hints, e.g., IETF Manufacturer Usage Description (MUD), to identify IoT
devices. While these methods are powerful and accurate, they do not scale with
new IoT vendors and devices which are constantly added to the market.\\

\vspace{-0.75em}
\noindent{\bf Analysis of Passive Data.}
Signatures derived by instrumented testbeds have been leveraged to infer the
presence of IoT devices within homes and enterprises using data from a
residential ISP and an IXP~\cite{IMC2020-IoT}. Machine learning techniques
have been used to generate IoT signatures and infer their presence in traffic
flows of IoT
devices~\cite{TMC19_SivanathanGLRWVS19,LCN19_SivanathanGS19,trimananda2020packet}.
Recent work by Perdisci et al.~\cite{iotfinder} leverages
distributed passive DNS data collections combined with machine learning to
identify a variety of IoT devices based on their DNS fingerprints.
User agents have also been used to infer IoT devices in network
flows~\cite{IMC19_DekovenRMABSSVS19} or server logs~\cite{IMC2020-UAs}. Feng
et al.~\cite{feng2018acquisitional} uses machine learning to label IoT devices based on
information extracted from websites (vendors, Wikipedia, product reviews). Yu
et al.~\cite{yu2020you} propose to use deep learning to identify mobile and IoT
devices as well as their manufacturer and model by extracting features from
structural and textual information embedded in passively observed broadcast and
multicast packets from public WiFi networks. All of the above work focuses
on identifying IoT devices.\\

\vspace{-0.75em}
\noindent{\bf Active Scanning.} 
Izhikevich et al.~\cite{LZR2021,IPv4Allports} perform active scanning campaigns and include IoT services that are often reachable on non-IoT ports. Kumar et al.~\cite{IoT-All-Things-2019} utilize data from an antivirus software that scans
home networks to discover IoT devices at home and assess their level of
security. This study---one of the largest of its kind---shows that IoT adoption
differs substantially across regions. It is widespread in North America, where
nearly half of the homes host at least one IoT device, typically an
Internet-connected television or streaming device. On the contrary, in South Asia, only around three percent of the homes host IoTs, typically surveillance
cameras. In total, the study discovered more than 83 million devices deployed
in roughly 16 million households. It also identified weak default credentials
and showed vulnerabilities to known attacks.

Companies periodically scan the Internet using a wide range of ports, including IoT standard ports, and offer annotated datasets~\cite{ZMap,censys,Shodan}.
Active measurement campaigns are used by, e.g., Srinivasa et
al.~\cite{open-for-hire} to detect IoT clients in the wild and characterize IoT
device misconfigurations. Note that they explicitly look for devices and not IoT
backends. The same is true for work that tries to identify IoT devices that
participate in attacks, e.g., the Mirai
attack~\cite{USENIXSS17_Antonakakis}.  %
We conclude that most work using active scans focuses on IoT devices and their
security properties.\\

\vspace{-0.75em}
\noindent{\bf IoT Platforms.} 
We are aware of a small number of studies that focus on IoT platforms, whereby
their main focus is also on security.  Alrawi et al.~\cite{alrawi2019sok}
perform a security evaluation of home-based IoT deployments and highlight the
need to understand IoT platforms, i.e., IoT backend providers. He
et~al.~\cite{He-iot-cloud} develop fingerprinting techniques to classify traffic that is exchanged with the cloud as IoT-related or non-IoT-related traffic. A
study by Zhou~et~al.~\cite{Zhou-iot-cloud} investigates five popular IoT
platforms that enable smart home IoT applications. The study shows that these
platforms are vulnerable to a number of removal attacks, including device
substitution, device hijacking, device denial of service, illegal device
occupation, and firmware theft. Jia~et~al.~\cite{Jia-iot-cloud} report on the
vulnerabilities of defense mechanisms used by popular IoT platforms for
IoT-specific protocols, e.g., MQTT. We note that none of the prior works
characterizes IoT backend provider footprints or their traffic flows.

\section{Methodology}\label{sec:methodology}

\begin{figure}[t]
	\captionsetup{skip=.25em,font=small}
	\includegraphics[width=1\linewidth,valign=t]{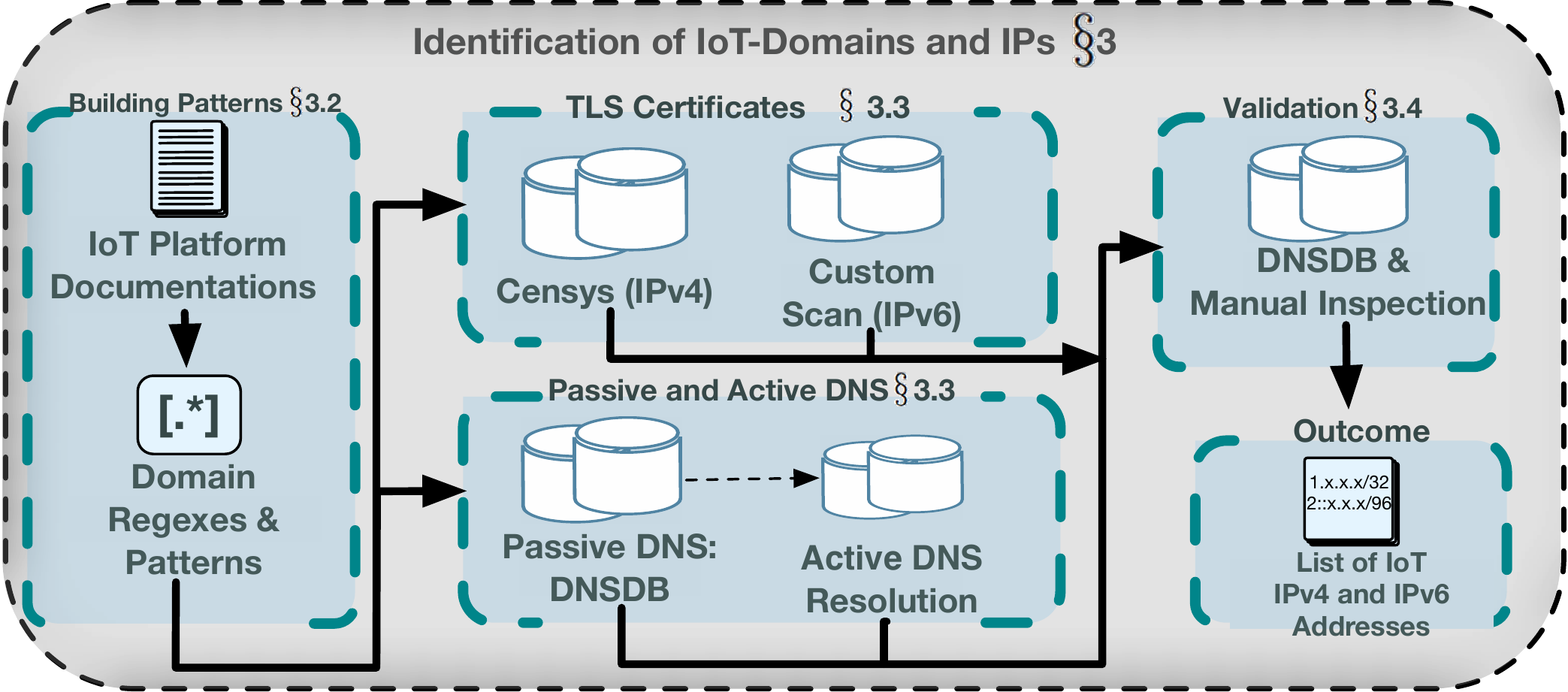}
	\caption{Our methodology to infer IoT backends' footprint.}
	\label{fig:hitlist-methodology}
\end{figure}

In this section, we discuss how we use a diverse set of sources, including
documentation by IoT providers, active and passive DNS measurements, and
IPv4/IPv6 scans to identify the set of backend IPs of each IoT backend
provider, see Figure~\ref{fig:hitlist-methodology}.

\subsection{Selection of IoT Backend Providers and Study Periods}

To compile a list of IoT backend providers, we consider the popular ones that
were mentioned in previous studies~\cite{Jia-iot-cloud}. We also expand this
list by considering IoT backends operated by other major manufacturers, e.g.,
Cisco, Huawei, and Siemens, and cloud providers, e.g., Oracle, Tencent, and SAP.
The complete list of the IoT backend providers in our study is presented
in~\Cref{table:summary}. By some accounts, for example, the IoT platform market
research report by IoT analytics~\cite{IoT-Analytics}, these IoT backend
providers are in the top 17 in terms of estimated revenue and are responsible
for more than 90\% of the total revenue.\footnote{Note that
these reports are not peer-reviewed and we use their reported IoT backend
revenue only for a rough estimation of their market share.
We use neither the ranking nor or the revenue of these companies in our methodology.}

We focus on the week starting from February 28 at midnight and ending on
March 7, 2022. We also collect preliminary results (only IPv4) for December
3 to 10, 2021. Since the results are consistent, we focus on the week
starting in February for all but Section~\ref{sec:aws-outage}.

\subsection{Identification of IoT Domain Patterns}
For each IoT backend provider, we start by identifying the domain names and the
IP prefixes that are used for the backend. This information is often contained in their publicly available documentation since IoT vendors and device programmers need it. When a backend provider explicitly discloses their IPs, we
use them for our validation (see Section~\ref{subsec:validation}). Typically,
IoT backend domains follow a well-defined form
{\tt<subdomain>.<region>.<second\-level-domain>},\\ 
where:

\begin{itemize}[leftmargin=*]
\item {\tt <subdomain>} is either a domain of a specific IoT services or a
  unique identifier (e.g., a hash or the name of an IoT platform customer). Some
  companies such as Alibaba, Tencent, and Bosch, also list the
  network protocol, e.g., MQTT, CoAP~\cite{alibabaiotdocs,tencentiotdocs,boschiot}.
\item {\tt <region>} indicates the full name or code of a city, a country, a region, or a
  continent;
\item {\tt <second-level-domain>} is either the second-level domain name of the parent company of the IoT backend provider or a special domain name
  allocated for the IoT backend.
\end{itemize}

However, some providers, e.g., Google, use the same fully qualified domain
names (FQDNs) for all of their customers. In such cases, we use these FQDNs.  In
Table~\ref{table:summary} we report the 16 IoT platforms for which we were able
to generate regular expressions for their IoT backend domain names using
their official documentation.

We leverage the structure of the IoT
backend domain names to generate the regular expressions. If the <subdomain> part of the IoT domain is a unique
value, e.g., a hash or a random string, we replace it with a regex
wildcard. Similarly, we replace the <region> part of the IoT domains with
appropriate regex terms that match the naming scheme of the different regions of
the provider. Note that we also obtain the naming schemes for the regions of the
providers from their documentation. Finally, we concatenate the regex terms with
the <second-level-domain> to create the regular expressions. See
Appendix~\ref{sec:appendix-regex} for examples.

\begin{table*}[!bpt]
\centering
{\small
\begin{tabularx}{\textwidth}{ p{3.4cm}|p{.4cm}|p{1.2cm}|p{.9cm}|p{1.2cm}|l|p{.6cm} } 
 \toprule
Backend Provider Name [Source]& \# AS & \# IPv4 /24 (IPv6 /56) & \# Lo\-ca\-tions & \# Countries & Protocols (Ports) & Stra\-te\-gy \\ 
\midrule
Alibaba IoT~\cite{alibabaiot,alibabaiotdocs,alibabaiotdocsipv6} & 2 & 73 (2) & 27 & 13 & MQTT(1883), HTTPS(443), CoAP(5682) & DI \\
Amazon IoT~\cite{awsiot, AWS-protocol, Amazon-regions-zones} & 4  & 9,000 (20)
& 18 & 15 {\tiny+Anycast} & MQTT(8883, 443), HTTPS(443, 8443) & DI \\
Baidu IoT~\cite{baidoiot, baidoiotdocs, Baidu-regions-zones} & 2 & 26 (1) & 2 & 1 & MQTT(1883, 1884, 443), HTTP(80, 443), CoAP(5682, 5683) & DI \\
Bosch IoT Hub~\cite{boschiot} & 1 & 290 (0) & 1 & 1 & MQTT(8883), HTTPS(443), AMQP(5671), CoAP(5684) & PR \\
Cisco Kinetic~\cite{Cisco-kinetic,ciscokineticdocs} & 2 & 14 (0) & 4 & 2 & MQTT(8883, 443), TCP(9123, 9124) & PR  \\
Fujitsu IoT~\cite{fujitsuiot} & 1 & 2 (0) & 2 & 1 & MQTT(8883), HTTPS(443) & DI \\
Google IoT core~\cite{googleiot, googlemqtt443} & 1 & 114 (11) & 77 & 14 & MQTT(8883,443), HTTPS(443) & DI \\
Huawei IoT~\cite{huaweiiot} & 1 & 26 (0) & 2 & 1 & MQTT(8883, 443), HTTPS(8943),  CoAP(NA) & DI \\
IBM IoT~\cite{ibmiot,ibmiotdocs} & 2 & 116 (0) & 12 & 8 & MQTT(8883, 1883), HTTP(S)(80,443) & DI\\
Microsoft Azure IoT Hub~\cite{azureiot, azureiotdoc} & 1 & 282 (0) & 39 & 16 & MQTT(8883), HTTPS(443), AQMP(5671), & DI\\
Oracle IoT~\cite{oracleiot,oracleiotproto} & 3 & 67 (0) & 10 & 8 & MQTT(8883), HTTPS(443) & DI+PR \\
PTC ThingWorx~\cite{ptciot} & 3 & 881 (0) & 10 & 8 & Protocol Agnostic & PR \\
SAP IoT~\cite{sapiot, sapiotdocs} & 6 & 2.929 (0) & 7 & 5 & MQTT(8883), HTTPS(443) & PR \\
Siemens Mindsphere~\cite{siemensmindsphere,siemensmindspheredoc} & 4 & 126 (1) &
3 & 3 {\tiny+Anycast} & MQTT(8883), HTTPS(443), OPC-UA & PR \\
Sierra Wireless~\cite{sierrawireless, sierrawirelessmqtt,sierrawirelessprivateapn} & 4 & 7 (2) & 4 & 4 & MQTT(8883,1883), HTTP(S)(80,443), CoAP(5682,5686) & PR \\
Tencent IoT~\cite{tencentiot,tencentiotdocs} & 5 & 47 (2) & 5 & 4 & MQTT(8883,1883), HTTP(S)(80,443), CoAP(5684) & DI \\
\bottomrule
\end{tabularx}
\caption{\small Selected IoT backends (alphabetical order) and their base
  characteristics for the study period, Feb.\ 28--Mar.\ 7, 2022. Dedicated
  Infrastructure (DI), Public Cloud Resources or CDN (PR). We plan to release the
  IoT domain patterns as well as the set of IPs. }
\label{table:summary}
}
\end{table*}

\subsection{Identification of Server IPs}
Next, we use the above regular expressions to identify the IPs of possible IoT
backend servers. Hereby, we rely on two complementary techniques. First, we
take advantage of the information available in TLS certificates. Second, we use
passive DNS data, namely, DNSDB~\cite{dnsdb,pDNS:2005}. Finally, we complement
the data with an additional active DNS dataset.

\smallskip
\noindent{\bf TLS Certificates.}
Censys~\cite{censys} continuously scans the IPv4 address space. In addition to
scanning for open ports across a wide range of port numbers, it performs
protocol-specific handshakes to collect banners; and it provides metadata, e.g.,
geolocation. These results are published on a daily basis. Motivated by the
previous results~\cite{SIGCOMM2021-hypergiants}, we use daily snapshots
matching our study period to identify certificates with domains that match our
regular expressions. The corresponding IPs are IoT backend provider IPs. Note,
we only use certificates~\cite{cangialosi2016measurement,chung2016measuring}
that are valid during the study period.

During our study period, Censys scans only IPv4 addresses.
To identify IPv6 addresses, we run 
active measurements using various IPv6 hitlists~\cite{gasser2018clusters}. Our
hitlists include IPv6 addresses that showed activity for popular
IoT ports,
i.e., 443 (HTTPS), 8883 (MQTT), 1883 (MQTT), and 5671 (AMQP).
We add support for these IoT protocols to
ZGrab2~\cite{zgrab2} and we use it to collect TLS certificates from these IPv6
addresses. We perform this data collection from a server located in Europe.
For a discussion on ethical considerations, we refer to
\Cref{sec:ethics}. 

\smallskip
\noindent{\bf DNS.}
We complement the above data with DNS data because scanning services typically
only download the default certificates. In some cases, scanning services may
not even be able to download the certificates, i.e., if the IoT backend
provider (e.g., Google) requires to supply the domain name via the Server Name
Indication (SNI) extension. In addition, other IoT backend providers, such as
Amazon, require the installation of a \emph{client certificate}, in particular,
for IoT protocols. In the absence of this certificate, the TLS handshake will fail.

DNS is another source of data for mapping domain names to IPs.
DNSDB is a passive DNS database that contains historical DNS queries and
replies for both IPv4 and IPv6 from multiple resolvers around the globe. We
choose DNSDB as it supports regular expressions and time-range queries.
For each IoT platform, we use DNSDB to collect all IPv4/IPv6 addresses in the
response for queries where $(i)$ the domain name matches the regular expressions
for the IoT platform, and $(ii)$ the query was issued within our study period.
In addition, during our study period, we also performed daily active DNS
resolutions for all domains identified via DNSDB (see ~\Cref{sec:ethics} for ethical considerations).
To perform these resolutions, we use three locations: two in Europe and one in the United
States.
Compared to a single location, using three vantage points increases our IP address coverage by $\approx$ 17\%.

\subsection{Validation of Server IPs}
\label{subsec:validation}
At this point, we have identified IPs related to IoT backend services. However,
we do not know if they are used exclusively for IoT services or if they also
host other services, e.g., Web services. In addition, we validate the
accuracy and coverage of discovered IP addresses against ground truth for three
IoT backend providers.

\vspace{1em}
\noindent \textbf{Shared vs.\ Dedicated IPs:} To identify IP addresses in our
candidate sets that also provide services unrelated to IoT, we use a methodology
similar to the one by Saidi et al.~\cite{IMC2020-IoT} and Iordanou et
al.~\cite{IMC18_IordanouSIL18}. For each candidate IP, we use DNSDB to identify
all the domain names that resolve to that particular IP. Next, we count the
number of domains that do not match the IoT domain pattern, but map to the
IP. If this count exceeds a threshold, we assume that it is not exclusively used
to offer IoT backend services. Through this process, we detect IoT backend
providers that use CDNs or host non-IoT services. While choosing the threshold,
we discover that Google uses two different sets of IPs: one
exclusively for IoT MQTT traffic and another for HTTPS traffic that is also
used for other Google services. In our IoT traffic flow analysis
(Section~\ref{sec:isp-view}) we focus only on those parts of the infrastructure that are exclusively used for IoT.

\vspace{1em}
\noindent \textbf{Validation Against Ground Truth:} While not all IoT backend providers publicly share their used IP ranges, three
of them do this at least partially. Our methodology
identified all the publicly listed IP addresses for Cisco and Siemens. Microsoft lists
network prefixes for its IoT backend service, which correspond to more
than 12,000 IPv4 addresses. Using our methodology, we identify 484 of these
IPs. All of them are within the listed prefixes. We conduct a study using
traffic data from a large European ISP, see Section~\ref{sec:isp-view}, and
check the traffic to the listed prefixes. We only identify 52 IPs that are
active. Out of these, our methodology misses only 4 IPs
which leads to an underestimation of the IoT traffic volume of less than 1\%.

\begin{figure}[t]
	\captionsetup{skip=.25em,font=small}
	\includegraphics[width=1\linewidth,valign=t]{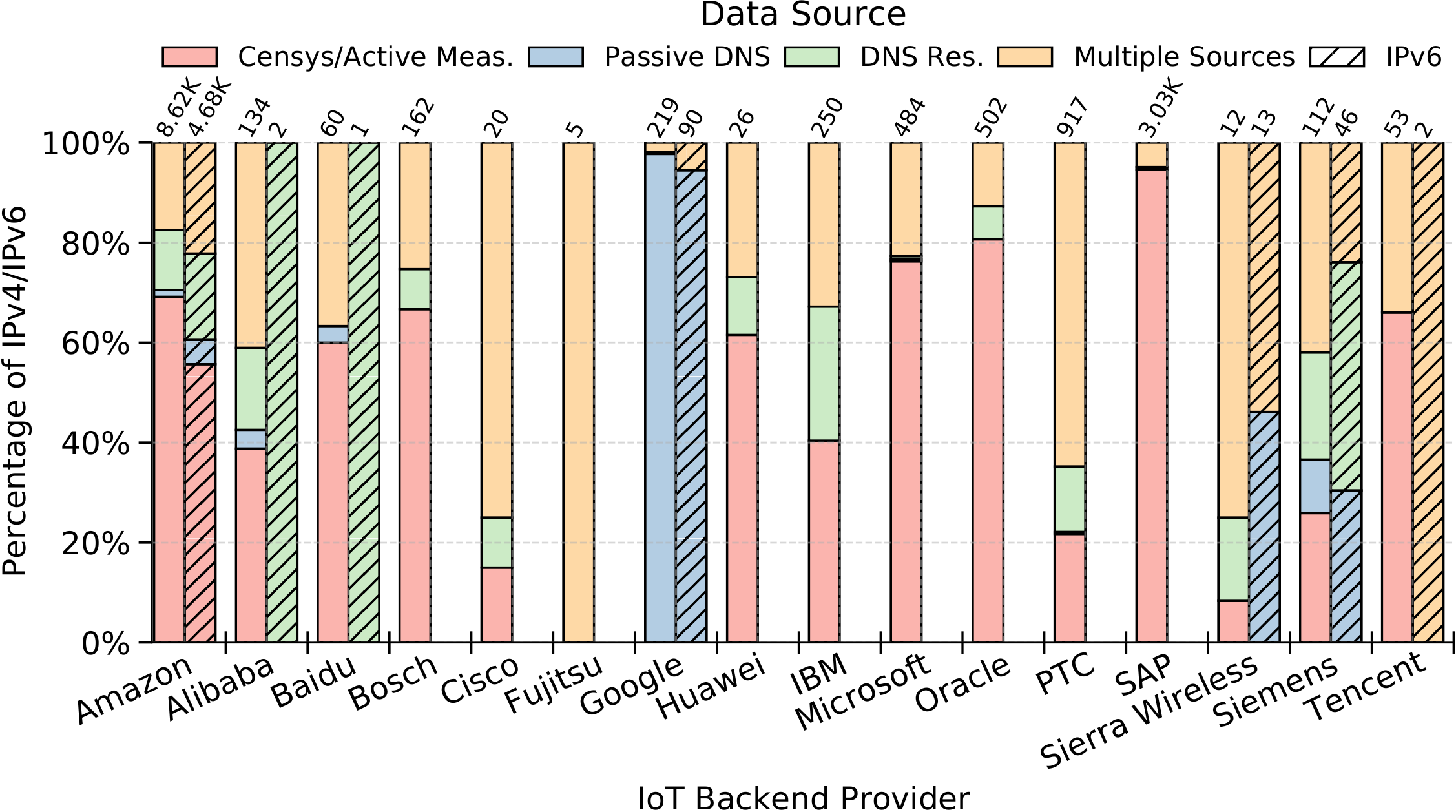}
	\caption{Fraction and \# of IPs per provider per source (left bar IPv4, right bar IPv6).}
	\label{fig:per-source-ip-fraction}
\end{figure}

\subsection{Contribution of Each Dataset}
Using as baseline the data collected on February 28, 2022, in
Figure~\ref{fig:per-source-ip-fraction} we show the contribution of each
data source grouped per IoT provider. The plot includes both IPv4 and IPv6
backends---the bar for IPv6 is shaded.
We distinguish IPs extracted from ``TLS Certificates'' (discovered via Censys),
our IPv6 scans, ``Passive DNS'' (discovered via DNSDB), ``Active DNS''
(identified via our active resolutions), and ``Multiple Sources'' (addresses
discovered by at least two methods).

First, we notice that some of the IoT backend providers only support IPv4
addresses. Second, there is no consistency regarding a preferred data source. For
example, when using only Censys data, we detect all IPs of the IoT backends for
Microsoft, SAP, and Tencent. But, we identify less than 2\% of the Google IPs. The
reason for this is that Google is using TLS SNI. Thus,
a majority of Google's IoT platform IPs are discovered using passive DNS. The contribution of
passive DNS is also substantial (more than 5\%) for Siemens, Alibaba, and Sierra
Wireless (IPv6). Our active DNS resolution is able to discover close to 20\% of
Alibaba (IPv4), Amazon AWS, Huawei, Bosch, Cisco, IBM, PTC, Siemens, and Sierra
Wireless, as well as the few Alibaba IPv6 server addresses. For the rest of this
paper, except if noted otherwise, we use the combined results of all techniques.

\subsection{Limitations}
The first limitation of our methodology relates to the stability of the IoT
domain patterns. IoT backend providers constantly update their service
infrastructure. This means that the patterns need to be regularly updated.
Moreover, not all providers publicly release their documentation. When the
documentation is not available, we do not try to identify those IPs due to
ethical concerns.

The second limitation is that some providers might not use TLS for their
services~\cite{alibabahttponly}. This might heavily impact the usefulness of
TLS scans, such as the Censys dataset. This limitation motivated us to augment
the scan data with DNS data. Still, even DNSDB has its own limitations, e.g.,
it does not have full coverage of all DNS requests.

Third, we leverage passive traffic data from an ISP in Europe to analyze IoT traffic in the wild.
Naturally, the vantage point's location might influence the overall IoT traffic that we see.

Finally, our ability to
discover IPv6 addresses is directly influenced by the coverage of the chosen
IPv6 hitlists \cite{gasser2018clusters,zirngibl2022rusty}.

\subsection{Ethical Considerations}\label{sec:ethics}

\smallskip
\noindent{\bf Active IPv6 Scanning.}
During the design and application of our methodology, we took care to
minimize any potential harm to the operation of routers and networks. First, the
load measurement is very low, i.e., a single packet per destination. We
also performed a randomized spread of load at each target IPv6 in the hitlist.
Moreover, we coordinated with our local network administrators to ensure that
our scanning did not harm the local or upstream network.

For the active scanning, we use best current
practices~\cite{ZMap,partridge2016ethical,dittrich2012menlo} to ensure that our
prober IP address has a meaningful DNS PTR record. We run a Web server with
experiment and opt-out information that responds to DNS resolution of the DNS
PTR domain. During our active experiments, we did not receive any complaints or
opt-out requests.  

\smallskip
\noindent{\bf DNS Resolution.}
We perform daily active DNS resolutions for all domains identified via DNSDB. We
make sure that the load in the DNS resolvers is low, i.e., we allow ten seconds
before subsequent resolution, and we utilize all the available resolvers. To
perform these resolutions, we used three locations, two in Europe and one in the United States. All the locations were well-connected to the Internet, and our
resolutions added negligible additional load to the network.

\smallskip
\noindent{\bf External Data.}
We applied for research accounts to both Censys and DNSDB. The accounts allowed
us to query and download the data that had been collected, i.e., active IP and
port scans, TLS certificates, and passive DNS requests and responses. We also
read the public documentation of IoT backend companies without using any
automatic download or web scraping tool.

\smallskip
\noindent{\bf ISP Data.}
The ISP's NetFlow setup explicitly captures header data only and no payload for
operational purposes.
The data is processed in situ and on the ISP's premise. Following best
operational practices, the NetFlow data is deleted at an expiration date set at
the data collection time. For our analysis, no data is copied, transferred, or
stored outside the dedicated servers that the ISP uses for NetFlow analysis.

Since parts of the NetFlow data can be used as Personal Identifiable
Information (PII) for subscriber lines they are anonymized. More specifically,
the data is anonymized by the BGP prefix before the data hits the disc. We also note that to minimize spoofing, the ISP uses best common practices, including network ingress filtering according to BCP 38~\cite{bcp38}.
To avoid \iotbackendprovider blocklisting and any leakage of information
related to traffic or the number of served subscriber lines, we agree to the terms for data analysis proposed by the European ISP, and therefore we anonymize the
names of all IoT backend providers when discussing ISP traffic.

\section{IoT Backend Characterization}\label{sec:characterization}
In this section, we provide insights regarding the deployment strategies of IoT backend providers for their Internet-facing gateways that enable the
communication between the IoT devices and the backend's internal systems.

\subsection{Stability of IoT Backends}

\begin{figure}
    \captionsetup{skip=.25em,font=small}
    \includegraphics[width=\linewidth]{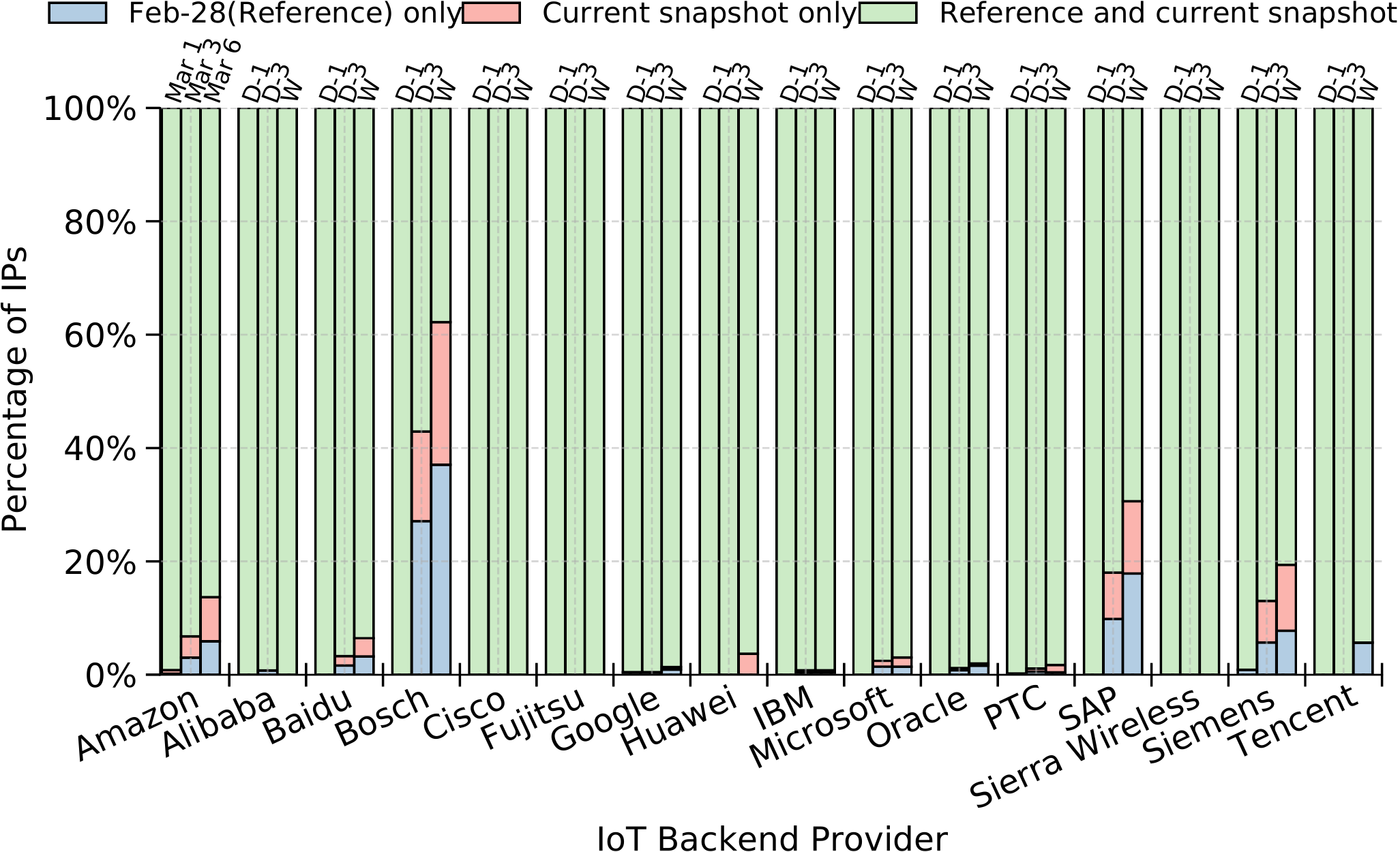}
    \caption{IoT backend: Stability of server IP set.}
    \label{fig:per-comp-stability}
\end{figure}

Before we dive into the characterization of the IoT backend deployments, we
evaluate how stable the set of discovered IoT backend server IPs---the
gateways---is across time. This gives us information on how frequently we have
to repeat our measurements. Using our weekly dataset,  in
Figure~\ref{fig:per-comp-stability}, we highlight changes in the daily IoT backend
server addresses per IoT backend. Our reference date is the first day,
February 28, 2022. The first bar for each backend compares it to the next day,
namely, March 1. We distinguish between IPs that are in both sets (green
bar), that are newly discovered (red), and those that are only in the first set
(blue). The other two bars are for March 3 and March 6.

We find hardly any change between the first two days. For most IoT backends,
there is also hardly any change within one week. This indicates that a weekly
measurement suffices. However, there are some exceptions, i.e., Amazon AWS,
Bosch, SAP, and Siemens. This is because these, at least partially, rely on shared public cloud infrastructure, as we show later. Their IP set is
more volatile, e.g., due to service scaling or service migration. However,
this is not necessarily the case as some cloud providers~\cite{elastic-ip}
offer static IPs. As such, the IoT backend IP usage also depends on the IoT company strategy. We use all IPs discovered
during the weekly study period for the remainder of this section.

\subsection{Footprint}

For the following reasons, it may be important for an IoT backend provider to
have a presence in multiple physical locations. First, having a footprint in
multiple datacenters and points of presence (PoPs) minimizes the impact of outages, physical disasters, or
attacks on a subset of them~\cite{espresso-google}. Second, datacenters from different regions are
useful for coping with regional demands and can improve application
performance. Third, it is increasingly important that datacenters for IoT
backends are available in different regions to comply with regulations regarding
transferring, processing, and storing data. For example, in the EU, the General
Data Protection Regulation (GDPR) poses constraints regarding data leaving EU
borders.

We use a number of heuristics to infer the footprint of each IoT backend. Many of the IoT backends, e.g., Google~\cite{google-regions-locations} and Baidu~\cite{Baidu-regions-zones}, encode the location in the domain
name. Typically they use either city level, e.g., two or three letters, or airport
codes. Others, e.g., Amazon~\cite{Amazon-regions-zones},
Alibaba~\cite{Alibaba-regions-zones}, and Huawei~\cite{huawei-cloud-locations}, use region codes in the domain name that can be mapped to cities using their
documentation. Using such hints, we are able to determine the footprint of all
IoT backends, except Oracle and a small subset of IPs. For these, we use
multiple sources, including the location of prefix announcements from Hurricane Electric, Censys geolocation information, and pings from traceroute looking glasses to locate each IoT backend server IP. Typically, all alternatives point
to the same location. In %
less than 7\% of cases, these sources report different locations, in which case we use the majority vote.

The results, see Table~\ref{table:summary}, show that the large majority of the IoT
backend providers use multiple locations in at least two countries. However,
there are exceptions: Baidu's and Huawei's backends are located only in
China. This is surprising given that Baidu and Huawei operate datacenters around the
world. Still, our extensive analysis allows no other conclusion.

Bosch offers a diverse range of IoT-related products, including machine
learning, data analysis, and device management. These components rely on
multiple public cloud providers in multiple locations around the globe, and each
component has to be purchased individually. The Bosch IoT Hub component is the only
one that offers a \frontend for IoT devices. Therefore, we restrict our study
to the locations and affiliated servers of the Bosch IoT hub, and it has a single
location.

IoT backend providers use different deployment strategies ranging from using
{\it Dedicated Infrastructure (DI)} to {\it Public Cloud Resources (PR).}
We say that an IoT backend uses DI if all its identified IP addresses are
announced by an Autonomous System that is managed by the backend. If the IP
addresses are announced by a cloud provider or CDN, we refer to it as PR. Of
our sixteen IoT backend providers, nine rely on dedicated infrastructure while six
rely on public cloud providers. Bosch IoT Hub, Cisco Kinetic, and Sierra Wireless on Amazon Web Services (AWS). 
PTC relies on the AWS and Microsoft clouds.
SAP IoT and Siemens Mindsphere rely on AWS, Microsoft, as well as Alibaba. 
Such diversity enables providers to improve their footprint and offer services in
many regions around the globe. The last IoT backend provider---Oracle---expands
his own dedicated infrastructure by leasing resources from Akamai (we label
this as DI+PR).

\subsection{Network Diversity}

First of all, we notice that the use of IPv6 is relatively low. We
discover IPv6 IoT backend server addresses for only seven of the 16 IoT backend
providers. Hereby, Alibaba offers IPv6 only in China, and Microsoft explicitly
states in its documentation that it does not yet support IPv6. Overall, the
number of discovered addresses is substantially smaller for IPv6 than IPv4, see
Table~\ref{table:summary}.

Network diversity, i.e., reachability of IoT backends via multiple ASes or
prefix diversity, is important to circumvent congestion, blocking, and network
misconfiguration, to enable fast reroute, and improve performance. We use
the RouteViews Prefix to AS mapping dataset from CAIDA~\cite{routeviews} to map
IP addresses to prefixes and AS numbers. Our analysis shows that all
IoT backend providers in our study use multiple, in some cases tens of prefix
advertisements, typically from more than one AS. Thus, we can expect that
short-term routing or availability disruption lead to minor service
degradations. Indeed, given the many available IPs and prefixes, it should be
possible to use DNS to redirect IoT requests to available and well-performing IoT backend
servers. In Section~\ref{sec:outage}, we revisit this hypothesis when studying a large-scale outage 
of one of the largest cloud providers.

Six IoT backend providers, Bosch, PTC, Siemens, SAP, Sierra Wireless, and Cisco, rely on one or more public cloud providers. This enables them to cope
with the short-term unavailability of outages. Also, as mentioned earlier, Oracle
uses its own dedicated infrastructure as well as that of a CDN. At least two IoT backend providers, Amazon IoT and Siemens,  also use
\emph{anycast} or, more specifically, the Amazon Global Accelerator
service~\cite{awsaccelerator}. Anycast services aim to map IoT
requests servers close to the client and cope with disruptions.
This highlights that IoT backend providers care about reliability and diversity.

\subsection{Protocol Support}

In Table~\ref{table:summary}, we also report---per IoT backend provider---the
supported protocols as listed in their documentation. They all claim to
support MQTT, an often used protocol for IoT messaging. The protocol is 
lightweight, follows the publish-subscribe paradigm, and is designed for machine-to-machine
communication. However, the IoT backend providers use different MQTT
ports. Some use the default unencrypted MQTT port 1883. The majority uses the
encrypted MQTT port 8883. Other providers also use non-standard
ports, i.e., non-IANA assigned to a protocol. For example,
for MQTT %
Baidu listens on port 1884. 
At least three IoT backends, i.e., Amazon, Baidu, and Google, use the secure Web port 443 for MQTT.

In addition, they often offer support for other IoT-specific protocols,
including CoAP and AMQP. The ports vary, e.g., include 5682 and 5684 for
CoAP. Baidu supports CoAP requests on multiple ports, i.e., 5682 and 5683. AMQP
is the least popular protocol among our IoT backend providers and is offered on
port 5671. We also observe some application-specific protocols, e.g., Siemens
offers OPC-UA, while PTC offers a protocol-agnostic communication platform.
The majority of the IoT backend providers also support Web protocols, namely
HTTP on port 80 and/or HTTPS on port 443.

We conclude that IoT backend providers quite often use non-expected ports. Thus, purely probing the expected ports can be misleading. This is
in line with recent results that observed unexpected applications running on
servers~\cite{LZR2021}. The motivation for offering different ports
even for the same IoT protocol, e.g., MQTT, may be to circumvent port
blocking. This is likely the reason why MQTT service is offered on port
443 by some of the providers~\cite{googlemqtt443}.

\section{IoT Traffic Flows}\label{sec:isp-view}

So far, we have used our methodology to understand the footprint of the IoT backends. Next, we use traffic information from a large European Internet Service Provider (ISP) to study IoT traffic patterns.

\subsection{Vantage Point}

Our vantage point is a major European ISP offering residential Internet IPv4
and IPv6 connectivity to more than fifteen million broadband subscriber
lines. The ISP uses NetFlow~\cite{Cisco-Netflow} to monitor the traffic flows 
at all border routers of its network, using a consistent sampling rate across
all routers. This data is needed to support daily operations as well as
network planning. For the ISP analysis, we anonymize all IoT company names (cf. ~\Cref{sec:ethics}).

\smallskip
\noindent{\bf Study Periods.}
For our IoT traffic flow analysis, we match the study periods for which we
identify the footprint of the IoT platform providers, i.e., February 28 to
March 7, 2022. In addition, we do a focused study during an outage,
see Section~\ref{sec:outage}, for December 3--10, 2021.

\begin{figure}[t]
    \captionsetup{skip=.25em,font=small}
    \includegraphics[width=\linewidth]{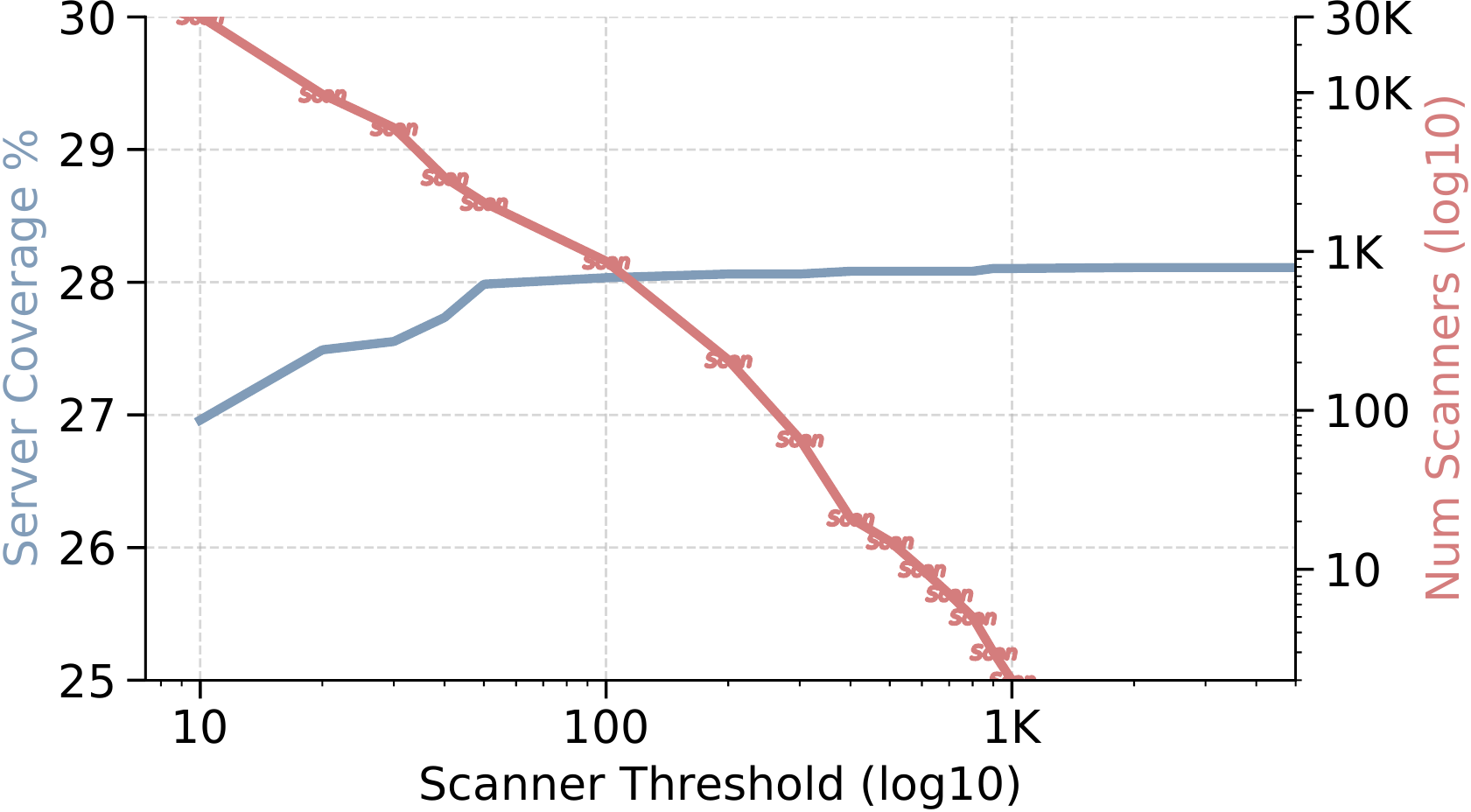}
    \caption{Scanner threshold vs.\ \% IPv4 IoT backends (blue line, left
      y-axis) and \# scanning subscriber lines (red line, right
      y-axis (log)).}
    \label{fig:scanner-threshold}
\end{figure}

\subsection{IoT Backend Platforms: Visibility}

Our characterization of \iotbackendproviders has shown that they often rely on
a global footprint to offer their services globally. Our first 
analysis, thus, focuses on the visibility of the IoT backend servers from our vantage
point, \ie the European residential ISP.

Our first validation check is whether any servers are within the address
space of the residential ISP. This applies to none, which is 
expected as we study the traffic of subscriber lines.
Our next check is for IoT backend infrastructure visibility from our vantage point, i.e., which fraction of the identified backend server IPs are contacted by subscriber lines of the ISP. Hereby, we do not expect that all servers are contacted as traffic localization, and other operational criteria within the IoT backends should map the ISP subscriber lines to a subset of their servers.

\smallskip
\noindent{\bf Exclusion of Scanners---Global Visibility.}
However, before proceeding, we have to exclude potential scanners within the
ISP since their scan traffic may bias our estimation of the visible part of the
IoT backend infrastructure. Scanners typically scan all or a substantial
fraction of all IPv4 IPs, resp.\ IPv6 IPs of the IPv6 hitlist. Therefore, a
subscriber line with a scanner is expected to send traffic to all IoT backend
servers. Therefore, we exclude scanners from our analysis which is possible as
the ISP uses spoofing prevention according to BCP 38~\cite{bcp38}.

To identify scanners, we follow the method proposed by Richter et
al.~\cite{scanningscannerphilipp}.
For each day during our study period, we compute the fraction of IoT backend server IPs that a subscriber line contacts. A subscriber line is said to host a scanner if it contacts more than a threshold of the server IPs.
Figure~\ref{fig:scanner-threshold} shows the results both for server coverage as well as ISP subscriber lines with scanners for February 28, 2022. More
precisely, we show how this fraction changes as we increase the strictness of
our criteria for identifying scanners---the scanner threshold (x-axis).
Hereby, our minimum scanner threshold is 10 IoT backend server IPs---a very strict selection criteria. We see that as we increase the scanner threshold, the number of scanners (red line and
right y-axis) decreases substantially. Yet, the percentage of IoT backend
servers that are visible does not increase drastically (blue line and left
y-axis).

\begin{figure}[t]
    \captionsetup{skip=.25em,font=small}
    \includegraphics[width=\linewidth]{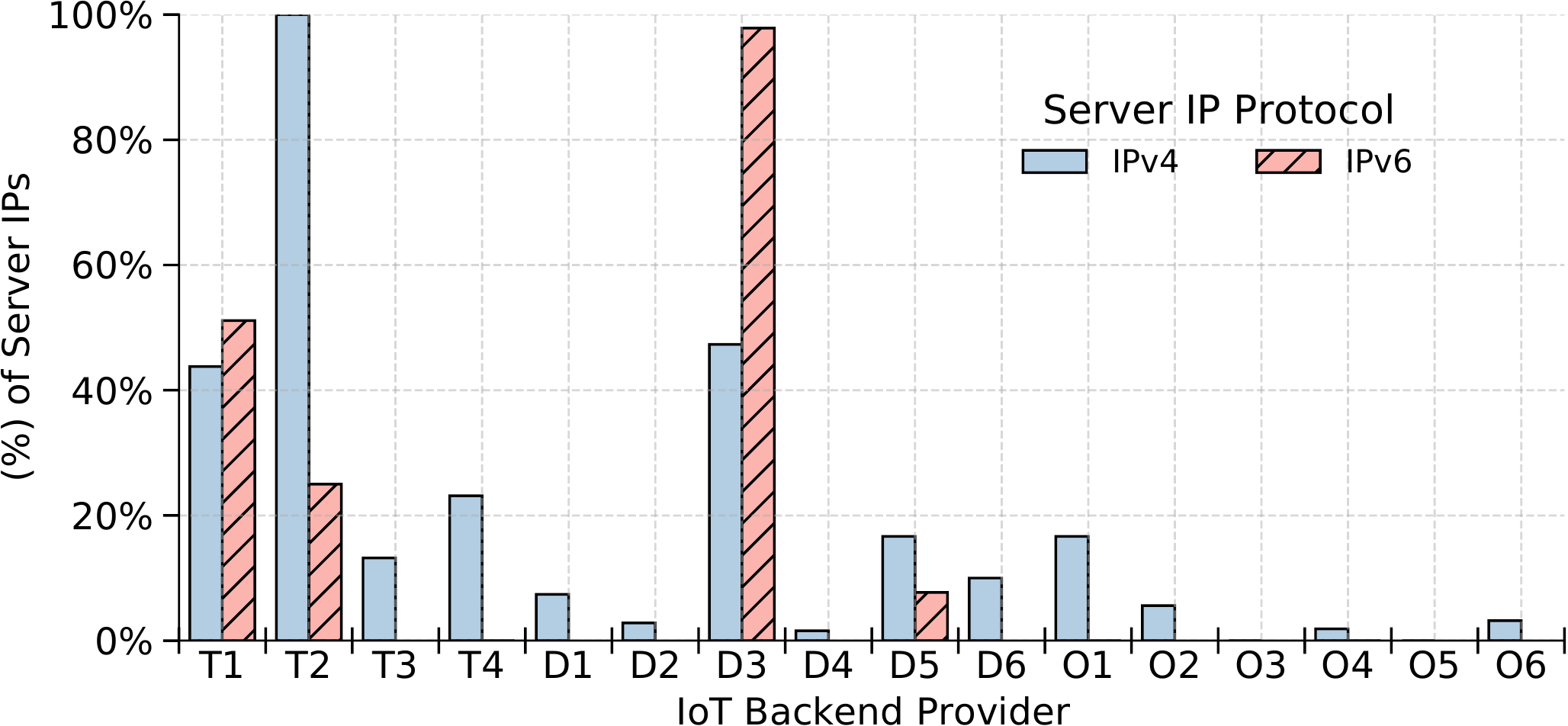}
    \caption{ISP vantage point: \% of Server IPs per IoT backend platform
      (Scanner threshold 100).} 
    \label{fig:servers-visible}
\end{figure}

We consider some baseline numbers: with a scanner threshold of 10 roughly 27\% of all identified IoT backend servers are visible while removing about 30k
subscriber lines. Using a threshold of 100 leads to the removal of less than 800 subscriber lines per day while resulting in a visibility of IoT backend
servers of approximately 28\%. As households often deploy multiple IoT devices
contacting 10 backend IoT IPs is still reasonable, as underlined by the large
number of subscriber lines. However, 100 server IPs are unlikely. As such, for
the rest of the paper, we use a scanner threshold of 100, which results in a
daily visibility of roughly 28\% of the identified IoT backend server IPs for
IPv4 and 51\% for IPv6 during our study period. Using this data, we
identify more than 2.32 million IPv4 and 202k IPv6 ISP subscriber lines with IoT activity per day.

\smallskip
\noindent{\bf Visibility per IoT Backend Provider.}
Next, we investigate if the visibility of IoT backend server IPs is uniform
across IoT platform providers. In Figure~\ref{fig:servers-visible}, we plot the
percentage of visible servers for each platform for IPv4 as well as
IPv6. As expected, the visibility varies substantially across the IoT backend
providers. For most, it is relatively small, between 5\% to 20\%. As remote IoT
backend servers should not be contacted by subscriber lines from a
European residential ISP, this is to be expected. Recall our insights from
Section~\ref{sec:borders} about the locations of the discovered IoT server IPs.
Surprisingly, for two \iotbackendproviders, namely \tone and \dthree, we
observe around half of the discovered IoT backend server IPs. Moreover, for one
IoT backend provider, namely \ttwo, almost all IoT backend server IPs are
visible. This provider is also among the top-4 popular \providers. On the other
hand for two other platform providers, namely \ofive and \othree, we hardly
find any activity. Since they are not focusing on the European residential
market, we exclude them from our analysis in this section.

\begin{figure}[bpt] %
    \captionsetup{skip=.25em,font=small}
    \includegraphics[width=\linewidth]{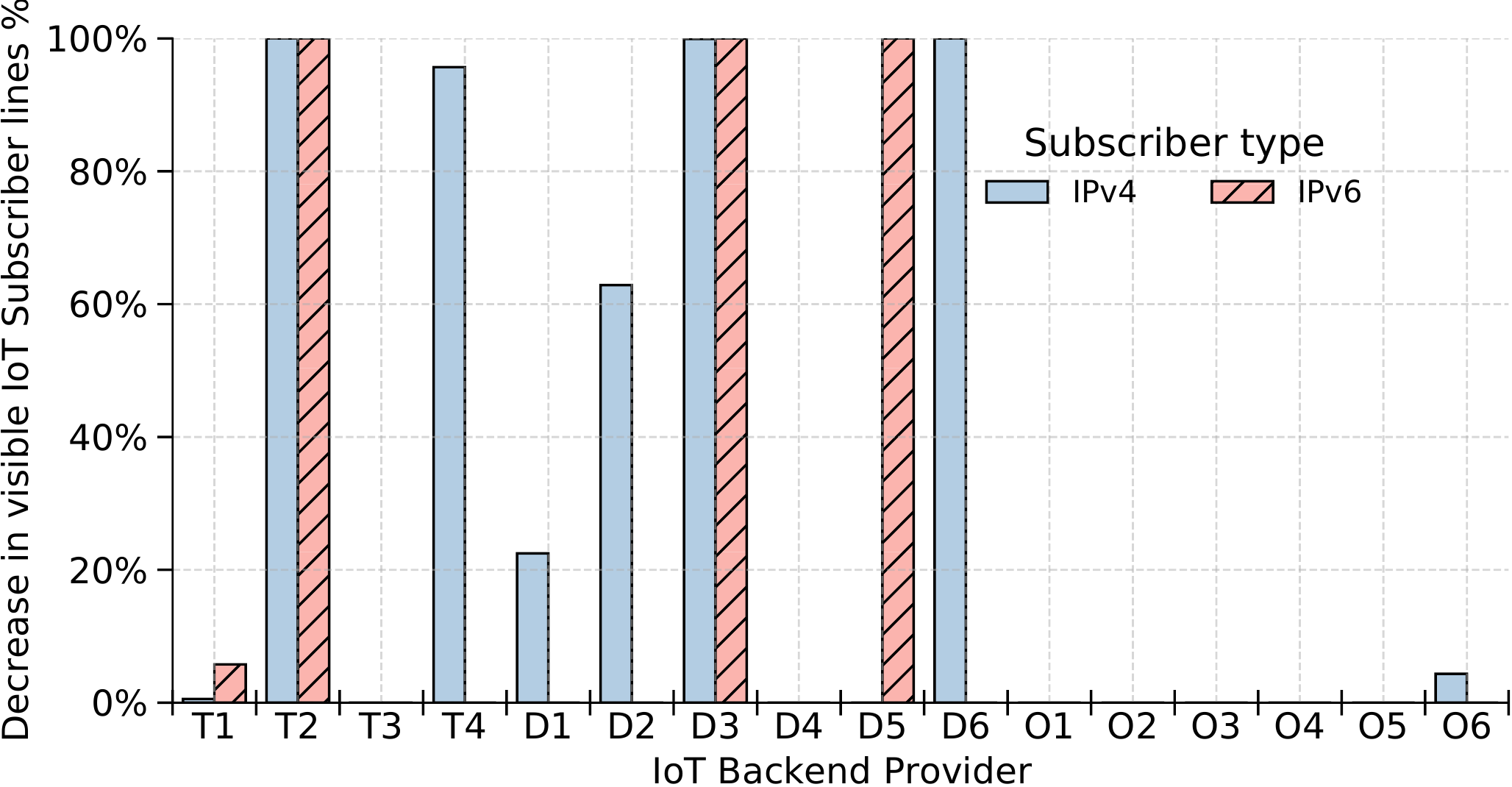}
    \caption{ISP vantage point---per IoT platform: \% decrease in ISP IoT
      subscriber lines by considering only TLS certificates.}
    \label{fig:per-comp-decrease-visible-clients}
\end{figure}

\begin{figure*}[bpt] %
    \captionsetup{skip=.25em,font=small}
    \includegraphics[width=\textwidth]{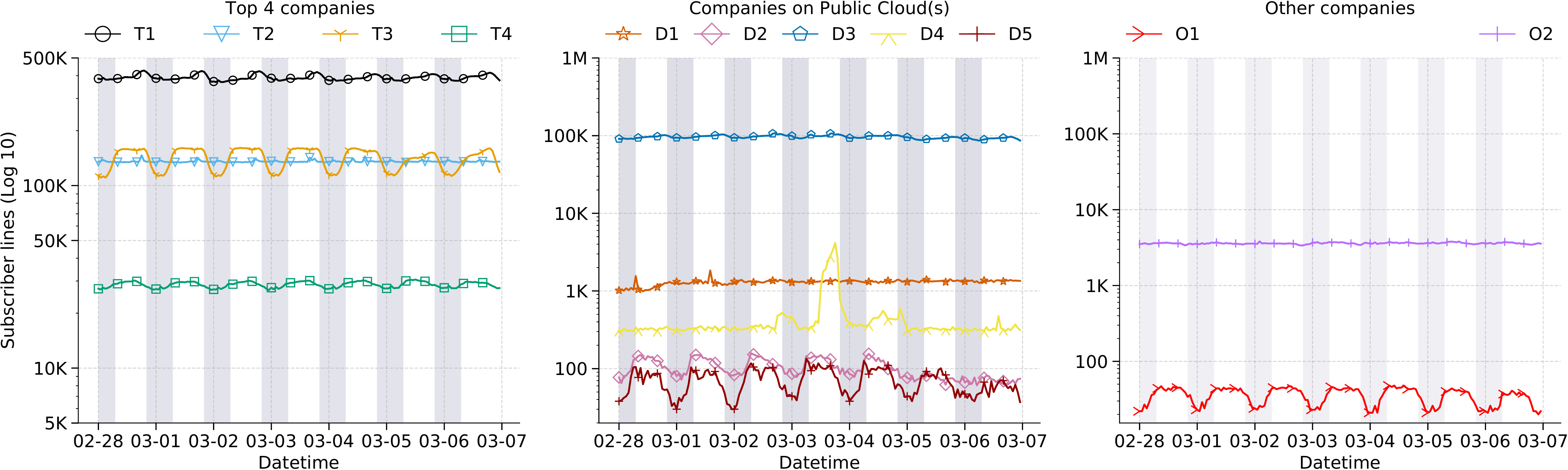}
    \caption{ISP vantage point---per IoT platform: \# of active subscriber lines.}
    \label{fig:per-comp-clients}
\end{figure*}

\begin{figure*}[t] %
    \captionsetup{skip=.25em,font=small}
    \includegraphics[width=\textwidth]{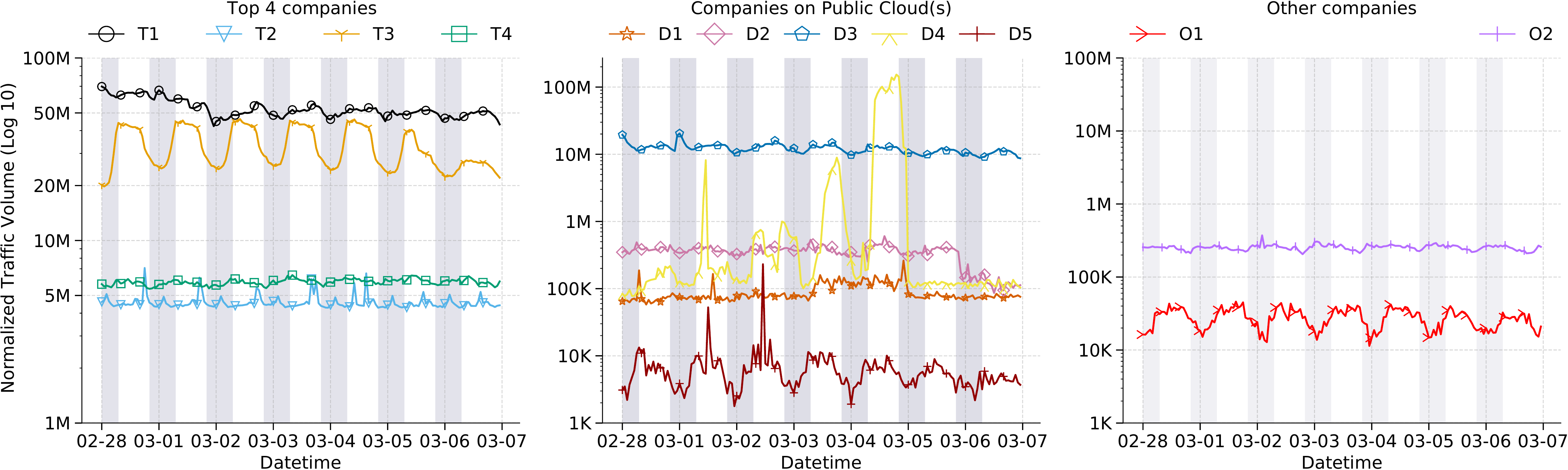}
    \caption{ISP vantage point---per IoT platform: Normalized total downstream
      traffic volume. }
    \label{fig:per-comp-bytes}  
\end{figure*}

\begin{figure*}[t] %
    \captionsetup{skip=.25em,font=small}
    \includegraphics[width=\textwidth]{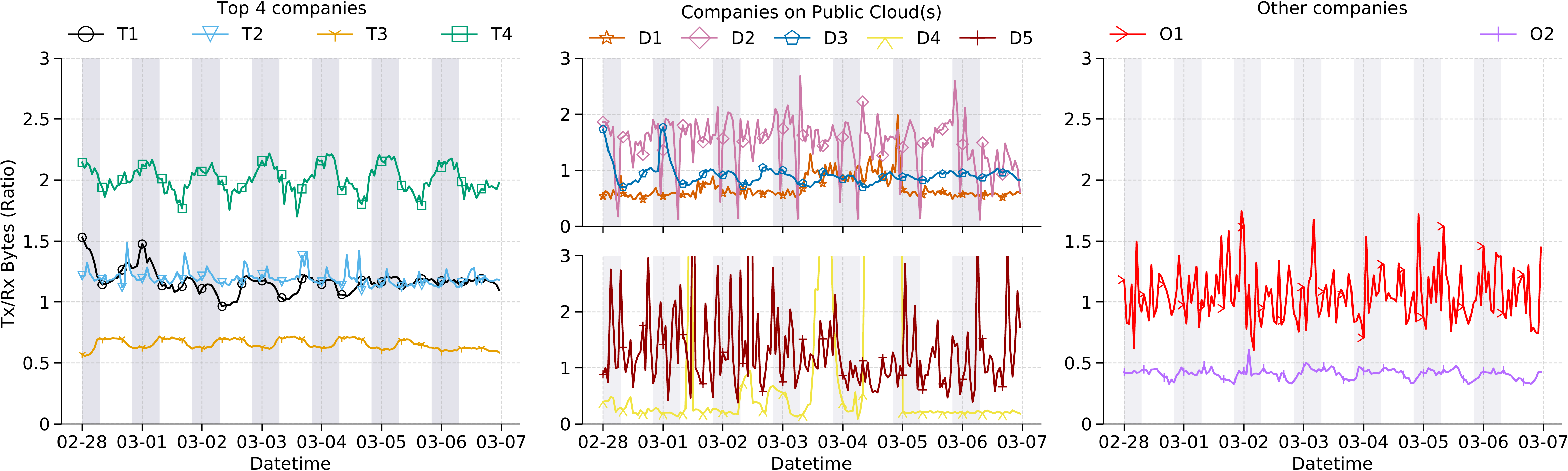}
    \caption{ISP vantage point---per IoT platform: Ratio of Downstream to
      Upstream traffic.} 
    \label{fig:per-comp-ratios}
\vspace{1em}
\end{figure*}

\subsection{ISP Subscriber Line Activity by IoT Backend Platform}

We find that a substantial fraction of ISP subscriber lines contact IoT
backend platforms. This underlines that the residential ISP is a suitable
vantage point.

\smallskip
\noindent{\bf ISP Subscriber Lines---Visibility by Data Source.}
While we know that our different data sources increase the discovery of IoT
platform server IPs, we do not yet know how important this is for discovering
IoT traffic. Thus, we check the necessity of using different data sources,
namely TLS certificates vs.\ passive and active DNS data. For this, we plot, in
Figure~\ref{fig:per-comp-decrease-visible-clients}, the decrease in discovered
subscriber lines with IoT traffic when we rely only on TLS certificate
information gathered by active IP scans (the Censys data set). For some IoT
platform providers, e.g., \tfour, \dsix, \ttwo, and \dthree, almost none
of the subscriber lines would have been detectable. Note that two of these are
providers that rely on SNI.

\smallskip
\noindent{\bf ISP Subscriber Lines---Activity across Time.}
Next, we explore how ISP subscriber line activity changes during our study, see
Figure~\ref{fig:per-comp-clients}.  It plots the hourly number of subscriber
lines for each \iotbackendprovider across the week. To plot the subscriber line
activity, we consider three subgroups of \iotbackendproviders, namely, the
top-4 per revenue, the ones that depend on cloud providers, and the remaining
ones. We only include those with at least 15 subscriber lines per hour. 

Figure~\ref{fig:per-comp-clients} (left) shows the activity of the top-4
\iotbackendproviders. We use light shading for the night---8 pm to 8 am local
time---to help in identifying the time of day effects. First, the level of
subscriber line activity differs substantially---in fact, by orders of
magnitude. Some have a clear diurnal pattern, e.g., \tthree, while others,
e.g., \ttwo, are more or less constant. We also observe that the peak time
differs among these \iotbackendproviders. The peak time for \tone and
\tfour is during prime time, i.e., between 6--10 pm, while for \tthree, it is
constant during the day, i.e., between 8 am and 8 pm. We attribute this to the
type of services that IoT devices offer and how often they communicate with
their \iotbackendproviders. For example, some IoT devices are likely to
be used at home for entertainment during prime time, while others offer
services that are used at any point in time.

Next, we move to those \iotbackendproviders that rely on the public clouds, see
Figure~\ref{fig:per-comp-clients} (center). Again, we see a large difference in
their usage across the board. Moreover, their activity does not correlate to
the one of the platform providers (plot not shown). Similar observations hold
for the remaining \iotbackendproviders, see Figure~\ref{fig:per-comp-clients}
(right).

\subsection{IoT Backend Traffic}

Next, we look at traffic levels. Here, we observe similar patterns as in the
IoT subscriber lines analysis which is expected as many of the IoT
applications are triggered by subscriber lines activity. 

\smallskip
\noindent{\bf IoT Backend Traffic---Downstream Volume.}
We find that the relative traffic volume level changes substantially, see
Figure~\ref{fig:per-comp-bytes}. It shows the normalized downstream traffic
volume for the same groups of \iotbackendproviders as before, namely, top-4,
public cloud dependent, and others. We notice that the traffic volume per
subscriber line differs substantially. On the one hand, even though the number
of observed subscriber lines differs by an order of magnitude for \tone and
\tthree their total traffic levels are relatively close to each other. On
the other hand, even though \ttwo and \tthree are serving a similar number of subscriber lines, their traffic volume differs by more than a magnitude. The
reason for this is that the traffic demands of IoT devices depend on the
applications. Thus, we conclude that the number of subscriber lines served by
\iotbackendproviders is not a good indicator for the downstream traffic
volume level of the \provider. This holds for all \iotbackendproviders that
we study.

\smallskip
\noindent{\bf IoT Backend Provider Traffic---Traffic Ratio.}
We also notice that the downstream and the upstream traffic demands of IoT
applications differ. Some of the IoT applications are heavy upstream, e.g.,
video surveillance, while others are heavy-downstream, e.g., online media
streaming. This is reflected in the IoT backend traffic. In
Figure~\ref{fig:per-comp-ratios}, we plot the ratio of downstream vs.\ upstream
traffic for the \iotbackendproviders of our study. Values above 1 indicate that
the \iotbackendprovider sends more traffic to the IoT devices than it
receives. Our analysis highlights that \iotbackendproviders differ. In all
three groups, namely, top-4, public cloud dependent, and the rest, we can find
heavy downstream as well as heavy upstream ones. Indeed, there is no particular
pattern to it. The ratios range from less than 0.33 to more than 3, which shows
that there is substantial asymmetry in the downstream vs.\ upstream ratio.
Moreover, we do not notice correlations between the ratios and the number of observed subscriber lines nor the downstream traffic level.

\begin{figure}[!tb]
    \captionsetup{skip=.25em,font=small}
    \includegraphics[width=\linewidth]{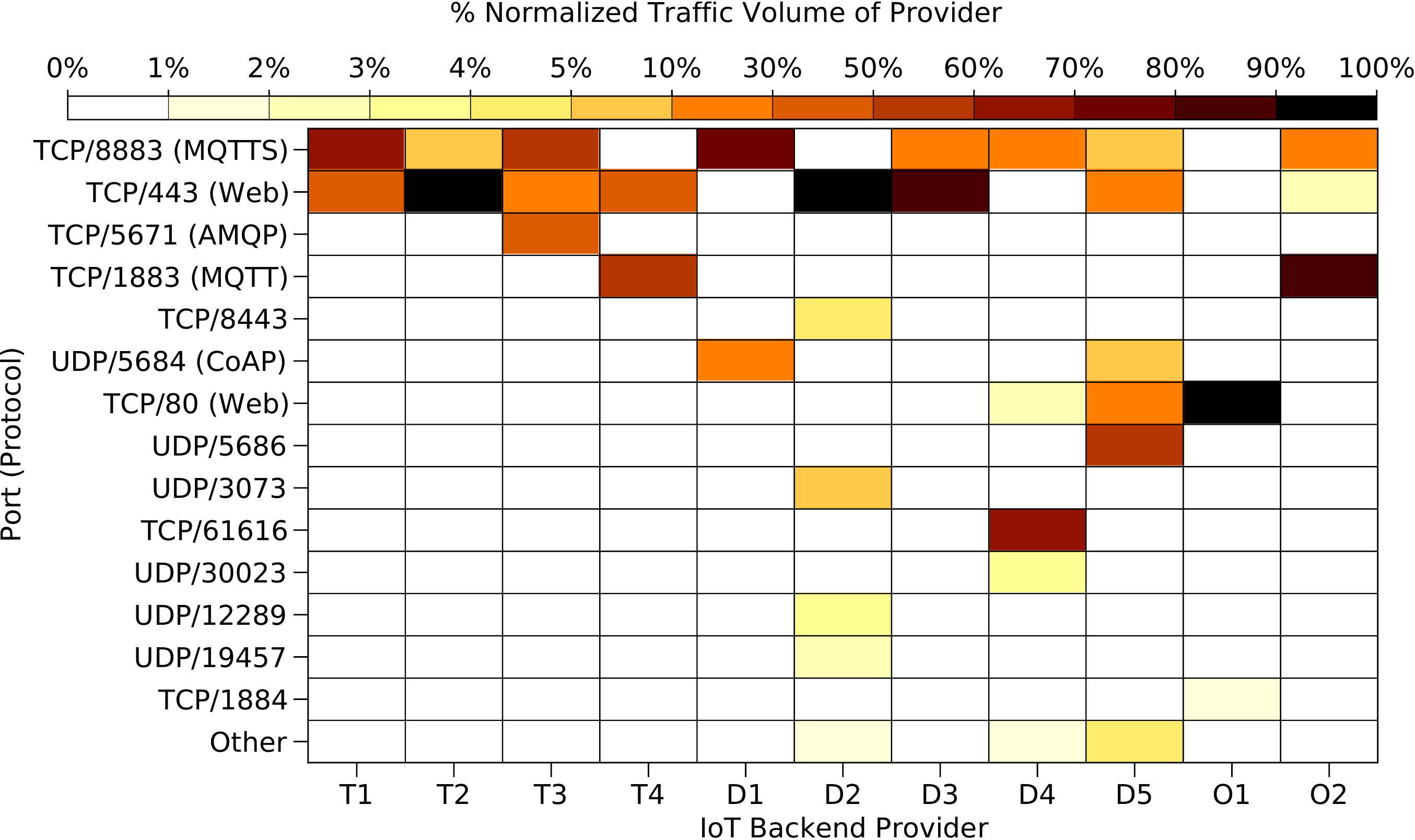}
    \caption{ISP vantage point: \% traffic volume per port, IoT platform.}
    \label{fig:per-comp-port-clients}
\vspace{-.5em}
\end{figure}

\subsection{\iotbackendprovider---Port Usage}\label{sec:iot-bend-port}

Next, we explore which network ports the IoT devices are using. Are they
relying on general-purpose application layer network protocols such as HTTP or
HTTPS, or are they using IoT-specific application protocols? Accordingly,
Figure~\ref{fig:per-comp-port-clients} shows the application layer protocol
mix as identified by IANA assigned port numbers for each \iotbackendprovider,
i.e., the percentage of traffic for each application protocol. Again, there is
no single pattern that describes all \iotbackendproviders.

Many utilize the popular Web secure ports, e.g., 443, typically over TCP. Its
usage varies from 5\% up to 90\%. IoT-specific protocols, e.g., MQTT, are also
popular. However, which MQTT port is used differs across
\iotbackendproviders. IANA assigns port 1883 for the non-secure version and port 8883
for MQTT over TLS. However, recall that some \iotbackendproviders, as per their
documentations, also offer MQTT service over non-standard ports such as 1884 or
even 443. The reasons for serving MQTT over non-standard ports include
the reduction of attack surface by reducing discovery probability via scans
and circumvention of firewalls that block standard MQTT. 

\begin{figure*}[t]
    \centering
    \captionsetup{skip=.25em,font=small}
    \begin{subfigure}[t]{0.3\textwidth}
        \centering
        \includegraphics[width=\textwidth]{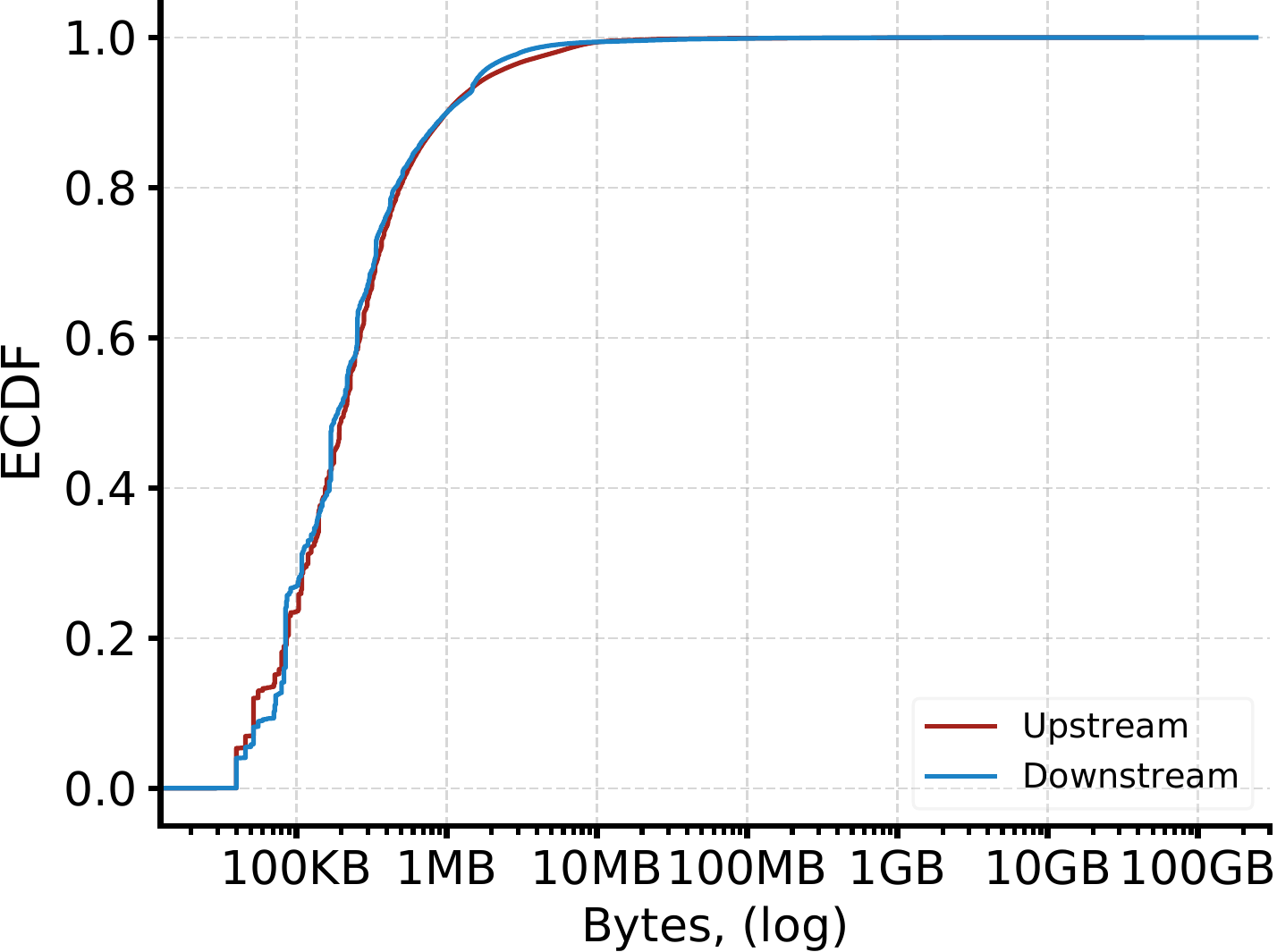}
        \caption{Estimated traffic exchanged in a day between a subscriber line and all
IoT backends in our study.} 
        \label{fig:per-comp-subscriber-bytes}  
    \end{subfigure}
    \hfill
    \begin{subfigure}[t]{0.3\textwidth}
        \centering
        \includegraphics[width=\textwidth]{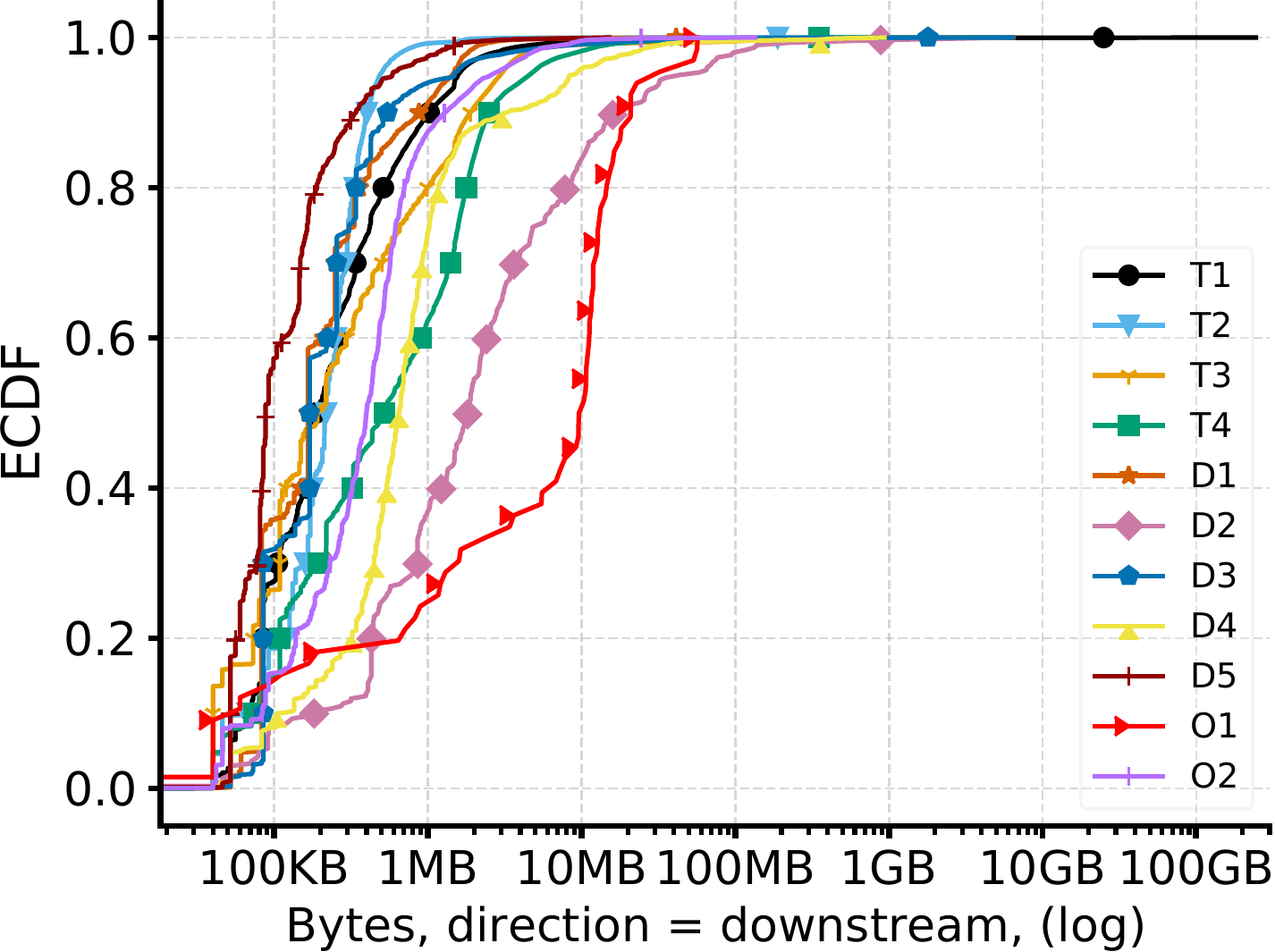}
        \caption{Estimated traffic exchanged in a day between a subscriber line and
each IoT backend in our study. In this plot, we consider the download traffic
per subscriber line.}
        \label{fig:per-comp-subscriber-bytes-separated}  
    \end{subfigure}
    \hfill
     \begin{subfigure}[t]{0.3\textwidth}
        \centering
        \includegraphics[width=\textwidth]{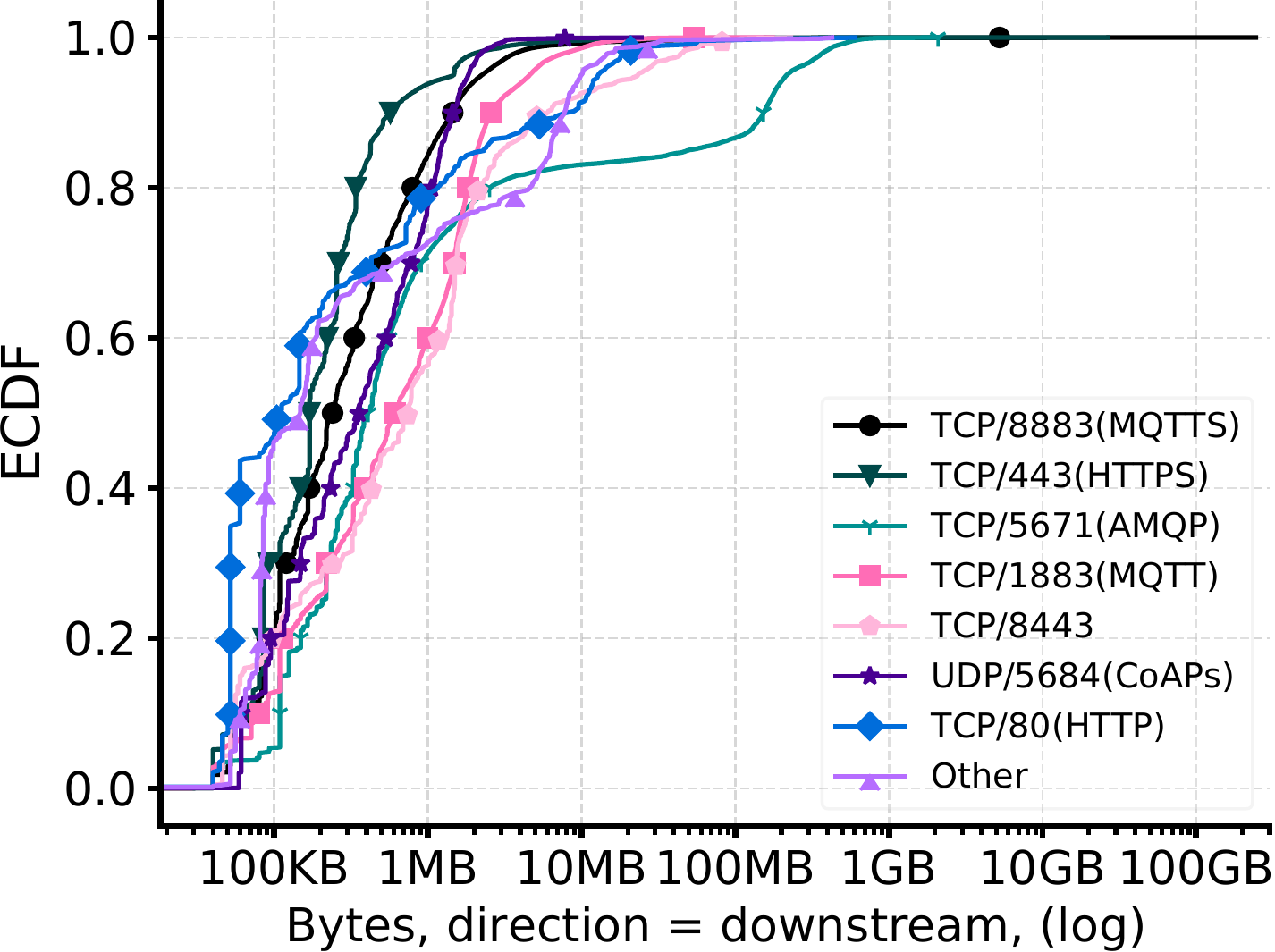}
        \caption{Estimated traffic exchanged in a day between a subscriber line
and all IoT backend providers for popular ports. In this plot, we consider the
download traffic per subscriber line.}
        \label{fig:per-comp-subscriber-bytes-port}  
    \end{subfigure}
    \caption{ISP vantage point: Traffic characteristics for traffic exchanged
in a day between a subscriber line and IoT backend providers or popular ports in
our study.}
    \label{fig:per-comp-subscriber-stats} 

\end{figure*}

We find that secure MQTT over its standard port is quite popular and used by
more than 50\% of all studied \iotbackendproviders. Other popular IoT
protocols include CoAP and AMQP. Similar to MQTT, some providers offer CoAP
over non-standard ports, e.g., the neighboring ports 5686 and 5682. For
UDP/5686, we do observe activity. For one provider, namely \dfour, we see that it
exchanges substantial traffic volume over port TCP/61616. This port number
is the default port number of the popular messaging software, Apache
ActiveMQ~\cite{apacheractivemq}, which processes messages sent via IoT-specific
protocols such as MQTT and AMQP. We further observe a number of UDP ports
above 10000 in use by various \iotbackendproviders.

Overall, this diverse port usage confirms previous insights~\cite{LZR2021} that
port scanning and protocol handshake do not suffice to uncover IoT backend
server infrastructure. In addition, to capture IoT-related protocols, it is not
sufficient to aggregate traffic of IoT-specific protocols, as this misses
a substantial part of the IoT traffic, e.g., the one served using HTTP(s) ports.

\subsection{Traffic Characteristics}\label{sec:traffic-chars}

We also investigate the characteristics of the traffic exchanged between
subscriber lines and IoT backend providers. This is an important test to
validate that this traffic is not generic Web traffic or video streaming of
popular applications that can be misinferred as IoT-related traffic.

In Figure~\ref{fig:per-comp-subscriber-bytes}, we plot the empirical cumulative
distribution function (ECDF) of the estimated traffic exchanged in a day between
a subscriber line and all the IoT backend providers we consider in our study. We estimate
the exchanged traffic considering the sampling rate. We plot both the download
and upload traffic exchanged, as some applications may be download-dominant or
upload-dominant. Our analysis shows that for the vast majority (more than 99\%)
of the subscriber lines, both the upload and download traffic exchanged with all
the IoT providers is less than 10 MB per day. This value is substantially lower than
the reported traffic consumed by smart TVs or residential users, which is no
less than 1 GB per
day~\cite{smartTV-trackers,varmarken2022fingerprintv,mazhar2020characterizing}.
Thus, we conclude that the traffic exchanged between subscriber lines and IoT
providers is unlikely to be general Web or popular application video traffic.

Then, we investigate if any of the IoT providers we consider in our study
deviates from the above mentioned behavior and offers general Web or popular
video streaming applications in the identified prefixes. In
Figure~\ref{fig:per-comp-subscriber-bytes-separated}, we plot the empirical CDF
for the estimated traffic exchanged in a day between a subscriber line and each
of the IoT backends we consider in our study for the download traffic volume.
Although there are differences across IoT providers, the general observation is
that the vast majority of the exchanged traffic is relatively low, i.e., less
than 10 MB per day. Thus, the IoT backend servers for each of the IoT
backend providers we consider in our study are unlikely to be used for general
Web or popular video traffic. Similar observations are made when we analyze the
upstream traffic.

Finally, we investigate if the traffic exchanged using specific ports indicates
the exchange of heavy traffic. In
Figure~\ref{fig:per-comp-subscriber-bytes-port} we plot the traffic exchanged
between subscriber lines and IoT backend providers for the most popular ports in
our study. We consider the downstream direction and the top-7 ports that
contribute to more than 95\% of the exchanged traffic and the aggregation of the
rest of the ports. Our analysis shows that there is only one port, namely, port
5671 (this port is registered with IANA for the secure version of the AMQP protocol), where
around 18\% of the subscriber lines exchange between 100 MB and 1 GB
per day. The high traffic volume exchanged is observed only in one of
the IoT providers, and it is a very small fraction of the overall traffic we observe
in our measurements. Similar observations are made when we analyze the upstream
traffic. We conclude that the vast majority of the traffic exchanged at
different ports between the subscriber lines and IoT backend providers do not resemble the general Web or popular video traffic. 
 
\subsection{Crossing Region Borders}\label{sec:borders}

Since the recent EU General Data Protection Regulation (GDPR) poses
restrictions on the transfer of data outside the EU, and since the transfer of data
to remote servers may impact the performance of delay-sensitive applications, we
next study how many of the European ISP's subscriber lines with IoTs contact
IoT servers outside of Europe. Hereby, we take advantage of the location
information collected for each IoT backend server IP.

\begin{figure}
    \captionsetup{skip=.25em,font=small}
    \includegraphics[width=\linewidth]{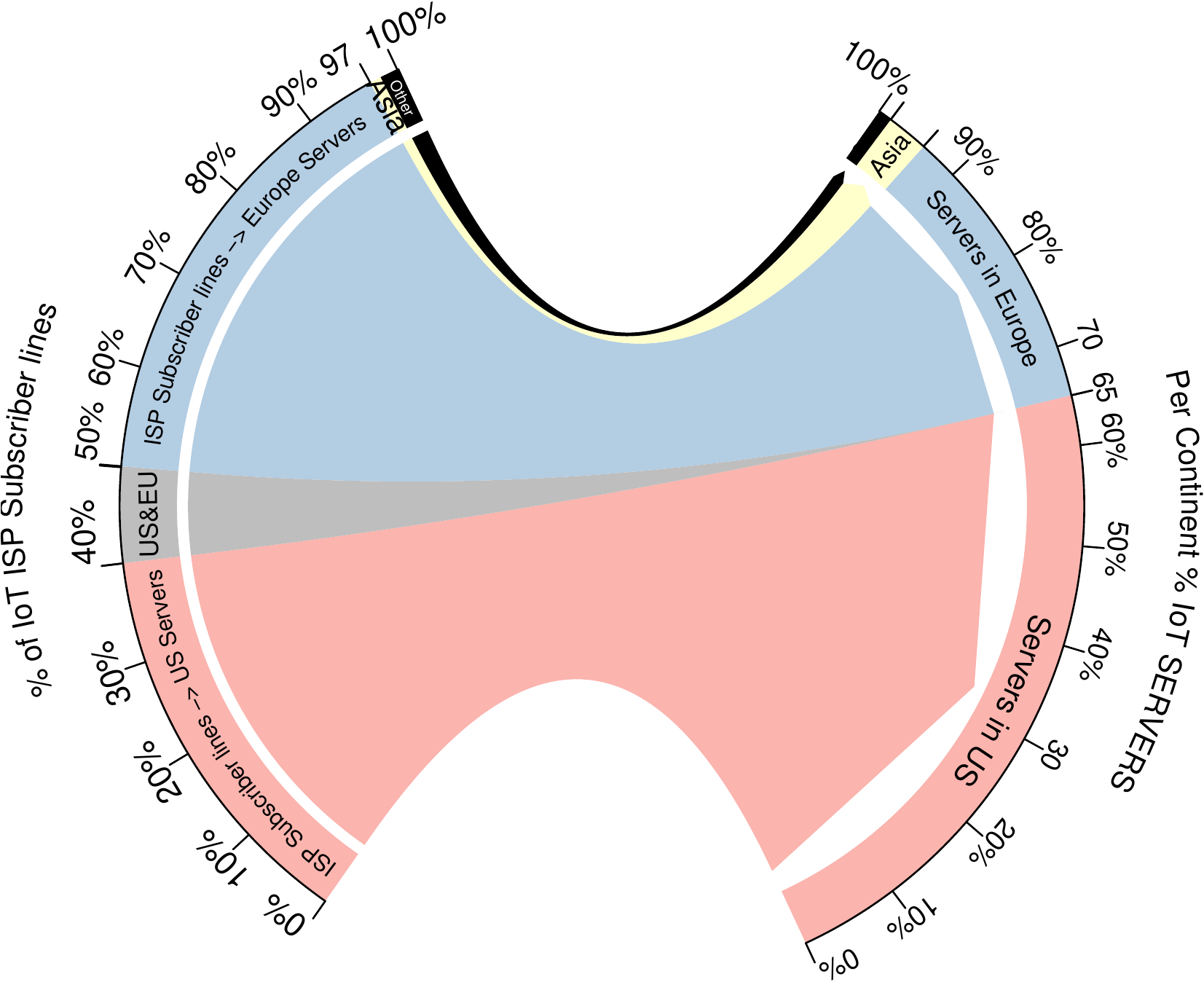}
    \caption{\% of ISP subscriber lines communicating with \% of Servers in each continent.}
    \label{fig:per-continent-clients}
\end{figure}

In Figure~\ref{fig:per-continent-clients} we visualize the percentage of
IoT-hosting subscriber lines, see left-side of plot, that exchange traffic with
IoT backend servers in different regions, namely, Europe, the US, Asia, and others.
Our analysis shows that slightly less than half, i.e., around 47\% of the
IoT-hosting subscriber lines communicate exclusively with IoT backend servers
located in Europe. Around 40\% of the IoT-hosting subscriber
lines contact IoT backend servers in the US. Around
10\% of the IoT-hosting subscriber lines contact a mix of 
locations from the EU and US. Around 3\% of the IoT-hosting subscriber
lines are contacting only IoT backend servers in Asia or other regions.

\begin{figure}   
    \captionsetup{skip=.25em,font=small}
    \includegraphics[width=\linewidth]{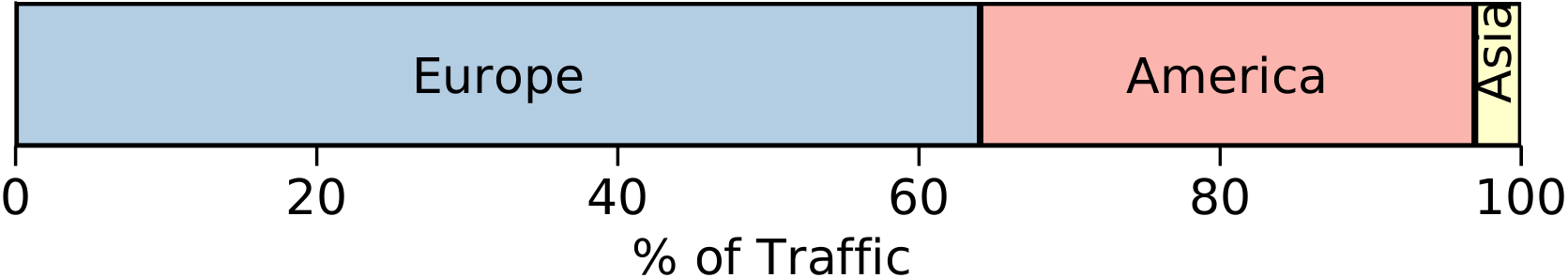}
    \caption{\% of ISP subscriber lines traffic exchanged with Servers in each continent.}
    \label{fig:per-continent-traffic}
 \end{figure}

On the right-hand side of the plot, we visualize the percentage of IoT servers
that are hosted per continent. We see that the IoT backend servers in Europe
are a minority of contacted servers, only around 30\%. Indeed, the majority of
the IoT backend servers, i.e., around 65\%, are located in the US. Around 5\% of the IoT backend servers are located in Asia, and a very small fraction elsewhere. We
conclude that around half of the IoT-hosting subscriber lines in the European
ISP contact IoT backend servers located in Europe, although they account for
less than one-third of the IoT backend servers identified in our study.

With regard to exchanged traffic volume between subscriber lines and IoT
backend servers, we notice that the majority of the traffic stays in Europe. In
Figure~\ref{fig:per-continent-traffic}, we plot the percentage of traffic
exchanged between subscriber lines and IoT backend servers annotated by
location. The largest traffic fraction, more than 62\%, is exchanged between
subscriber lines in Europe with servers in Europe. However, around 35\%---a
substantial fraction---is exchanged with servers in the US (where the majority
of the IoT backend servers are located). As such, IoT traffic is less localized
than one may have expected given the regulations of GDPR.

\section{IoT Backend Disruptions}\label{sec:outage}

In this section, we consider actual as well as potential disruptions to IoT backend providers. 

\subsection{AWS Outage}\label{sec:aws-outage}

During the time when we collected preliminary results (Dec.\ 3--10, 2021), a
major outage happened within the infrastructure of one of the major cloud
providers. More precisely, on December 7, 2021, Amazon Web Services, 
a cloud provider that is heavily used by the IoT backend providers we
study, experienced a large-scale
outage~\cite{aws-outage,aws-outage-fortune,aws-outage-te,aws-outage-wp} of its
US-East-1 service region (located in Northern Virginia). This outage affected
many popular websites and Internet services. Thus, we examine the effect of
this outage on the traffic flows of the IoT backend providers. 

\begin{figure} [!bpt]
	\captionsetup{skip=.25em,font=small}
        \includegraphics[width=1\linewidth,valign=t]{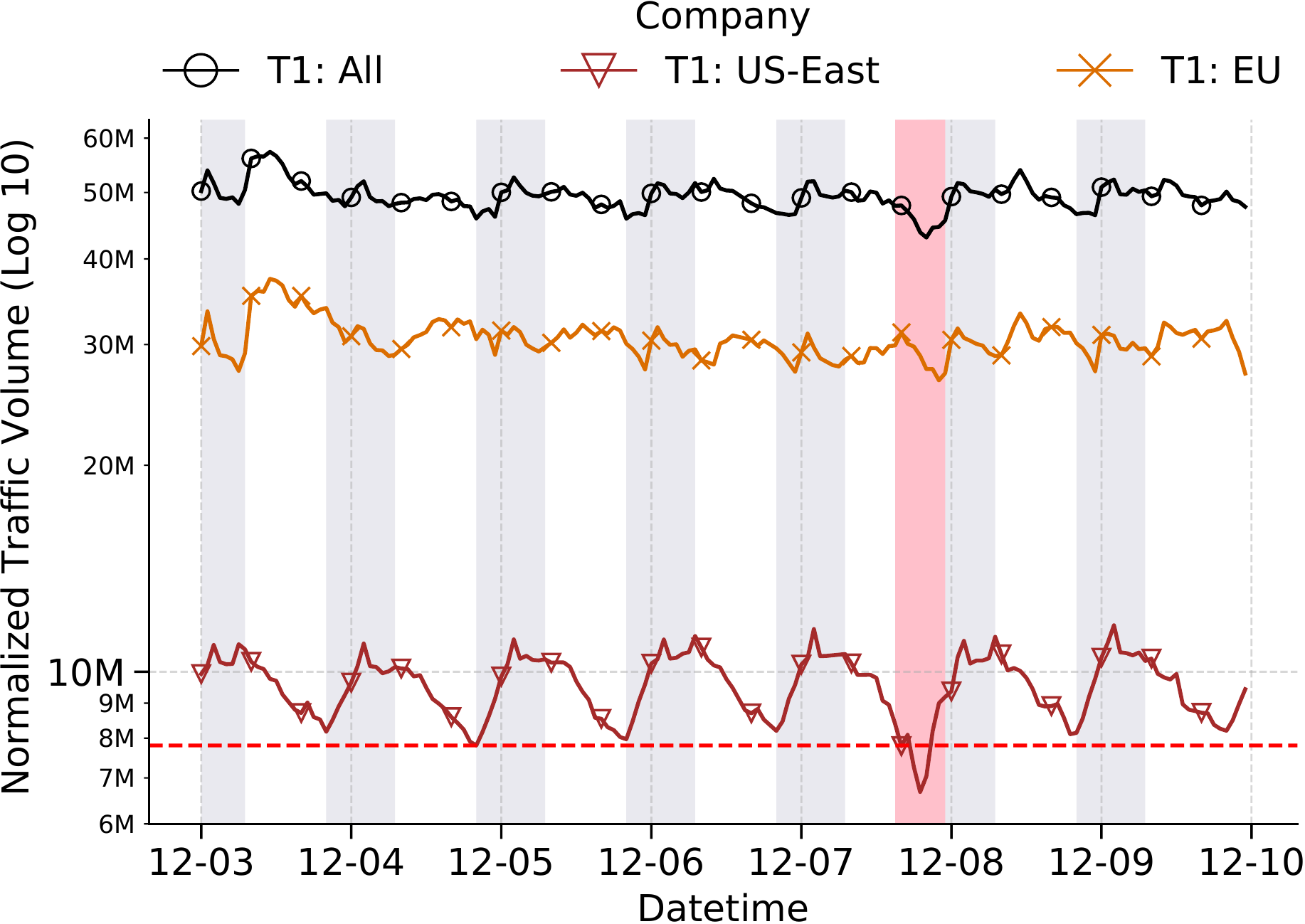}
	\caption{ISP vantage point---IoT backend provider T1: Normalized
          downstream traffic volume for all US east and EU service
          regions. The AWS outage is highlighted using a red
          background. The red line shows the normalized
          minimum traffic volume for the US east of the previous week.}
	\label{fig:aws-outage-bytes}
\end{figure}

\begin{figure} [!bpt]
	\captionsetup{skip=.25em,font=small}
	\includegraphics[width=1\linewidth,valign=t]{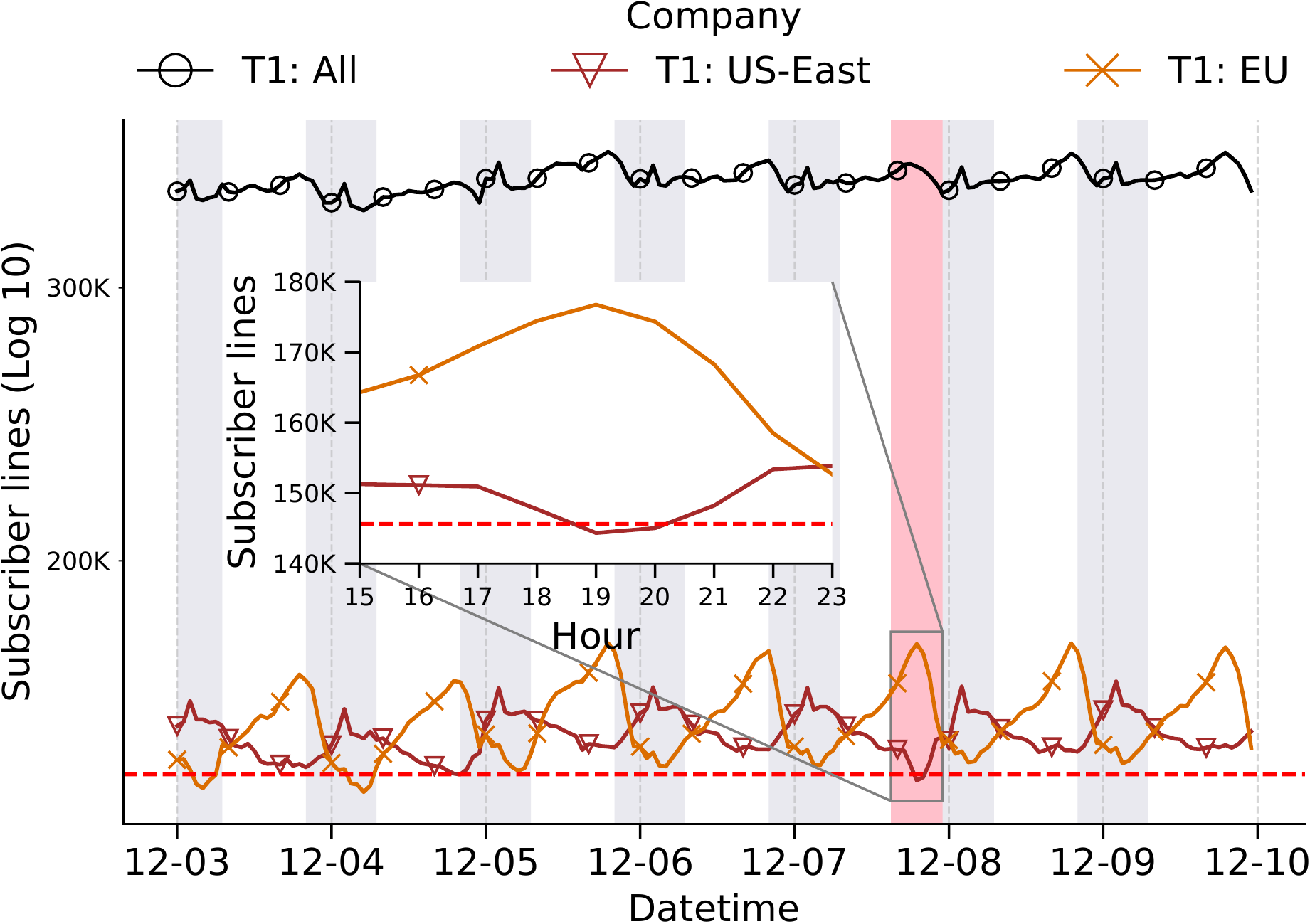}
	\caption{ISP vantage point---IoT backend provider T1: \# of subscriber
          lines for all US east and EU service regions. The AWS outage is
          highlighted using a red background. The red line shows the minimum \#
          of subscriber lines for the US east of the previous week.}
	\label{fig:aws-outage-clients}
\end{figure}

\smallskip
\noindent{\bf Impact on T1 ISP traffic flows.}
First, we analyze the outage's effect on the traffic from the \tone IoT
platform to the ISP's subscribers. Figure~\ref{fig:aws-outage-bytes} shows
\tone's normalized downstream traffic volume towards the ISP as well as the
normalized volume for two different AWS service regions, namely US east and
EU---aggregating the traffic of all US east resp.\ EU
availability zones. During the outage (highlighted with red background)
there is a substantial traffic drop---more than 14.5\% for the US east
coast region. Indeed, the total traffic volume is substantially lower than the
minimum observed traffic volume of the previous week (red line). This
highlights that cloud outages such as the one by AWS do not only affect Web
services~\cite{aws-outage-fortune,aws-outage-te,aws-outage-wp} but also IoT
services~\cite{aws-outage-bb,aws-outage-bi}.
When looking at the total traffic as well as the traffic from the EU sites, we
notice only slight dips. One reason is that the EU region services more than
three times more traffic than the US east coast region. Still, the fact that
there is a drop even for traffic in the EU region indicates some
interdependencies between the regions.  

\smallskip
\noindent{\bf Impact on T1 subscriber lines.}
Next, we check how these traffic volumes relate to the number of ISP subscriber
lines that contact IoT servers in these AWS regions, see
Figure~\ref{fig:aws-outage-clients}.  Again the red background highlights the
outage, and the horizontal line corresponds to the minimum number of subscriber
lines of the previous week. We see no impact for the EU region but a slight
decrease for the US East coast region. One may ask why this decrease is so
small. The reason is that we still observe the attempts of the IoT devices to
contact the servers in their assigned AWS regions. Thus, the downstream traffic
is lower, but the number of subscriber lines does not change drastically. Still,
it decreases, which indicates that some of them stopped trying, we did not observe them due to their decreased traffic volume or because they are remapped to
other regions.

\smallskip
\noindent{\bf Impact on \done--\dsix.}
Next, we explore if the outage also affected the IoT backend providers that
rely on AWS or the \tone IoT platform. We find hardly any effect, as the
subscriber lines of these platforms are mainly mapped to the EU AWS regions.

\subsection{Potential Disruptions}
\newcommand{\backendactiveipvfours}{16,203~}
\newcommand{\backendactiveipvfoursasns}{140~}
\newcommand{\backendactiveipvsix}{5,123~}
\newcommand{\backendactiveipvsixasns}{7}

Possible disruptions that we study for the week starting in Feb.\ 2022 are
connectivity problems due to routing and or IP filtering based on blocklists.

\smallskip
\noindent{\bf Connectivity problems.}
Such problems include routing problems such as BGP leaks, or BGP hijacks as well
as AS outages. We rely on Cisco's BGPStream service, which provides historical
information about BGP hijacks, leaks, and outages~\cite{bgpmon-cisco}. It
identified 10 BGP leaks, 40 possible BGP hijacks, and 166 AS outages. None of
these affected any of the identified IoT backend server IPs nor the ASes they
are hosted in.

\smallskip
\noindent{\bf IP Filtering.}
Next, we check how likely it is that a backend becomes unreachable as
a consequence of appearing in a blocklist. Here, we take advantage of the FireHOL
project\cite{firehol-project}, which generates a list of suspicious addresses,
by combining information from popular blocklists. In Feb.\ 2022, the FireHOL
blocklist contained over 610M IPv4 addresses extracted from 67
blocklists\footnote{We excluded one of the blocklists as it is known that it is
not carefully maintained, see
https://github.com/pushinginertia/ip-blacklist/issues/9, and is, thus, likely
to produce false positives.}.
Using daily blocklists matching our study period, we
check if the server IPs are included in any of the blocklists. We identified 16
such IPs. The non-exclusive reason for their inclusion in the blocklist are:
Four are associated with open-proxies and anonymizing services, one is 
linked to malware, and five are associated with network attacks/spam. Moreover,
nine originate from a personal
blocklist\footnote{https://graphiclineweb.wordpress.com/tech-notes/ip-blacklist/}.
These IPs belong to 6 of our IoT backend providers, namely, Baidu (5 IPs),
Microsoft (4 IPs), SAP (4 IPs), Google (3 IPs), Amazon (2 IPs), and Alibaba (1
IP).

\section{Concluding Remarks } %

IoT device population, as well as application complexity, have increased
substantially over the last decade. An ecosystem of IoT backend providers has been established to cope with the IoT-specific demand. Our study takes
advantage of the significant market consolidation~\cite{iot-consolidate}---less
than twenty IoT backend providers are responsible for more than 90\% of the
market share. These IoT backend providers are either IoT vendors or large cloud
providers offering services tailored to IoT developer needs. Our study focuses
on 16 IoT backend providers, including the top 10.

Discovery of the Internet-facing part of the IoT backends is a challenging task, as pure IP-port scanning misses a significant share of the addresses for many
IoT backend providers. Indeed, we find that the port usage differs substantially
across IoT providers. It is not unusual for IoT protocols, e.g., MQTT, to use
non-standard ports or to reuse Web ports. The latter makes the identification of IoT backend infrastructure as well as IoT traffic challenging. However, 
fusing data from publicly available documentation, certificate data from active
scanning, with passive and active DNS data allows us to unveil a detailed map of
IoT backend servers.

Our study shows that IoT backend providers' deployment strategies differ
substantially. While the footprints of most of them cover many geographical
regions, some are present in only one location. Yet, others are utilizing
anycast. Since this impacts service performance, it should impact IoT backend
provider selection. Moreover, regulatory compliance (e.g., GDPR and data
sovereignty compliance monitoring) related to IoT data
transfer, storage location, and processing also plays an increasingly important
role when selecting an appropriate IoT backend.
Surprisingly, around a third of the IoT traffic in our study is
exchanged with servers in different continents, although it could have been served from within the region of the IoTs. This raises questions regarding the configuration of applications and best practices when developing IoT applications, and it also raises questions regarding reliability. We find that a major outage of a cloud provider impacted some IoT services.

We also observe that six providers rely on another IoT backend provider
to expand their footprint or outsource IoT backend functionalities. %
Thus, outages that occur unexpectedly can have cascading effects.
For the one outage we studied in detail, this did not happen as these providers
used the regional service, which was not affected by the outage. Still, it is a
wake-up call to add flexibility and re-routing opportunities to handle 
IoT backend disruptions, e.g., outages, attacks, misconfigurations, blocklists.  

Our methodology also offers a scalable and lightweight approach to estimate
the popularity of IoTs and shed light on IoT application activity. This is
possible without the need to derive per IoT device/manufacturer signatures
using, e.g., instrumented testbeds. However, the methodology may also misclassify the traffic that goes beyond the IoT if the IoT backend providers reuse this infrastructure for other purposes and is not detectable by our DNS-based method. Moreover, the IoT backend providers' customers may also use this infrastructure, e.g., the MQTT servers, for non-IoT applications. Our traffic analysis highlights that 
the IoT population and activity per application differ vastly. While some
applications behave more like the typical user-generated traffic, i.e., diurnal
patterns, peak evening hours, and are downstream-heavy, this is not the case for
all IoT applications. In fact, some popular IoT applications' traffic peaks
during the day.

Looking ahead, we expect that the importance of IoT backend providers will
continue to increase as new IoT devices are constantly being added to the
Internet. As such, continuous monitoring of their footprint and related
traffic flows is crucial not just for compliance reasons but also to understand
how IoT is changing the Internet.
 
\section*{Acknowledgments}

We thank the anonymous reviewers and our shepherd Lorenzo De Carli for their constructive feedback. We are grateful to Censys~\cite{censys} and Farsight Security~\cite{farsight} for providing us research access to their datasets.
This work was supported in part by the European Research Council (ERC)
  Starting Grant ResolutioNet (ERC-StG-679158), the Atracción de Talento grant
  (Ref. 2020-T2/TIC-20184), funded by Madrid regional government, and by the SCUM
  Project (RTI2018-102043-B-I00) MCIN/AEI/10.13039/ 501100011033/ERDF.

\balance
\bibliographystyle{ACM-Reference-Format}
\bibliography{paper}

\begin{table*}[!bpt]

\centering
{\small
\begin{tabularx}{\textwidth}{l|l|l|l}
\toprule
Provider Name & Data Source & Api Type & Regular Expression/Query \\ 
\midrule
Huawei & DNSDB & Flexible Search & .$\backslash$.(iot-(coaps|mqtts|https|amqps|api|da)$\backslash$.).+$\backslash$.myhuaweicloud$\backslash$.com$\backslash$.\$/A \\ 
Amazon & DNSDB & Flexible Search & (.+)($\backslash$.iot$\backslash$.)([[:alnum:]]+(-[[:alnum:]]+)+)?($\backslash$.amazonaws$\backslash$.com$\backslash$.\$)/A \\ 
Oracle & DNSDB & Flexible Search & (.+$\backslash$.|$\wedge$)(iot$\backslash$.)([[:alnum:]]+(-[[:alnum:]]+)$\ast$$\backslash$.)?(oraclecloud$\backslash$.com$\backslash$.\$)/A \\ 
Baidu & DNSDB & Flexible Search & .$\backslash$.(iot$\backslash$.)([[:alnum:]]+(-[[:alnum:]]+)$\ast$$\backslash$.)?(baidubce$\backslash$.com$\backslash$.\$)/A \\ 
Siemens & DNSDB & Flexible Search & .($\backslash$.eu1$\backslash$.mindsphere$\backslash$.io$\backslash$.\$)/A \\ 
Sierra Wireless & DNSDB & Flexible Search & (.+$\backslash$.|$\land$)(na$\backslash$.airvantage$\backslash$.net$\backslash$.\$)/A \\ 
Bosch & DNSDB & Flexible Search & (.+$\backslash$.|$\land$)(bosch-iot-hub.com$\backslash$.\$)/A \\ 
IBM & DNSDB & Flexible Search & (.+$\backslash$.|$\land$)(internetofthings$\backslash$.ibmcloud.com$\backslash$.\$)/A \\ 
Microsoft & DNSDB & Flexible Search & (.+$\backslash$.|$\land$)(azure-devices$\backslash$.net$\backslash$.\$)/A \\ 
Tencent & DNSDB & Flexible Search & (.+$\backslash$.|$\land$)(tencentdevices$\backslash$.com$\backslash$.\$)/A \\ 
Tencent & DNSDB & Basic Search & rrset/name/$\ast$.tencentdevices.com./A \\ 
Google & DNSDB & Basic Search & rrset/name/mqtt.googleapis.com./A \\ 
Cisco & DNSDB & Basic Search & rrset/name/$\ast$.ciscokinetic.io./A \\ 
Amazon & Censys & String Search & $\ast$.iot.us-east-2.amazonaws.com \\ 
Amazon & Censys & String Search & $\ast$.iot.us-east-1.amazonaws.com \\ 
Amazon & Censys & String Search & $\ast$.iot.us-west-1.amazonaws.com \\ 
Amazon & Censys & String Search & $\ast$.iot.us-west-2.amazonaws.com \\ 
Huawei & Censys & String Search & $\ast$.iot-mqtts.cn-north-4.myhuaweicloud.com \\ 
Alibaba & Censys & String Search & $\ast$.iot-amqp.cn-shanghai.aliyuncs.com \\ 
Alibaba & Censys & String Search & $\ast$.iot-as-http.cn-shanghai.aliyuncs.com \\ 
SAP & Censys & String Search & $\ast$.iot.sap \\ 
\bottomrule
\end{tabularx}
\caption{\small An excerpt, less than 5\% of regular expresions and queries for a subset of IoT Backend providers.}
\label{table:regex}
}
\end{table*}
 
\appendix

\section{IoT Backend Regular Expressions}\label{sec:appendix-regex} 

This section provides an excerpt (\ie less than 5\%) of regular expressions for a subset of IoT
Backend providers in Table~\ref{table:regex}. For working with DNS records,
DNSDB offers two types of APIs, namely Flexible Search and Basic Search. A few sample queries that use our regular expressions are provided for each API type. For the full set of regular expressions, please see our released data~\cite{iot-backend-github}.

\end{document}